\documentclass[a4paper,11pt]{article}
\usepackage[english]{babel}
\usepackage{jheppub}
\numberwithin{equation}{section}

\newcommand{\tl}{\tilde{L}}

\newcommand{\rl}{\mathcal{L}}
\newcommand{\m}{\mathcal{M}}
\renewcommand{\r}{\mathcal{R}}
\renewcommand{\k}{\mathcal{K}}
\newcommand{\s}{\mathcal{S}}

\newcommand{\fk}{\mathfrak{K}}

\title{\boldmath Timelike Entanglement First Law and Linearized Field Equations in Higher Curvature Gravity}

\author{Mei-Hui Xiao,$^{a}$}
\emailAdd{xiaomh25@mail2.sysu.edu.cn}
\author{Guo-Ying Li,$^{a}$}
\emailAdd{ligy37@mail2.sysu.edu.cn}
\author{Song He,$^{b,1}$\note[1]{Corresponding author.}}
\emailAdd{hesong@nbu.edu.cn}
\author{and Jia-Rui Sun$^{a,2}$\note[2]{Corresponding author.}}
\emailAdd{sunjiarui@sysu.edu.cn}

\affiliation[a]{School of Physics and Astronomy, Sun Yat-Sen University, Guangzhou 510275, China}
\affiliation[b]{Institute of Fundamental Physics and Quantum Technology \& School of Physical Science and Technology,
Ningbo University, Ningbo, Zhejiang 315211, China}

\abstract{We investigate the timelike entanglement first law in holographic conformal field theories whose bulk dual is Lovelock gravity. Using the double Wick rotation formulation of timelike entanglement entropy together with the Jacobson-Myers entropy functional, we compute the linear variation of holographic timelike entanglement entropy for hyperbolic subregions. For cubic Lovelock gravity, we explicitly show that a single, universal multiplicative renormalization factor governs how higher curvature interactions enter the variations of both the entropy and the modular Hamiltonian, leading to $\Delta S=\Delta\langle H\rangle$ for low-energy thermal excitations. We then extend the analysis to Lovelock gravity of arbitrary order around the anti-de Sitter spacetime in the Fefferman-Graham gauge. For normalizable perturbations, the variation of the Jacobson-Myers functional reduces to the Einstein gravity's result multiplied by the same coupling-dependent factor that renormalizes the effective Newtonian constant in the linearized field equations of Lovelock gravity. We further show that the boundary contribution vanishes in the conformal limit for the class of perturbations considered. Consequently, the timelike entanglement first law is equivalent to the linearized field equations of Lovelock gravity about the maximally symmetric background, within the hyperbolic and perturbative regime considered in the present paper.}


\begin{document}

\maketitle
\flushbottom

\section{Introduction}
Entanglement, an essential concept of quantum theory, has played a crucial role in understanding quantum many-body systems and gravity~\cite{Calabrese:2004eu,Levin:2006zz,Kitaev:2005dm,Casini:2022rlv,VanRaamsdonk:2010pw,Maldacena:2013xja,Almheiri:2020cfm}. Typically, the measure of entanglement is called entanglement entropy, defined as von Neumann entropy $S_{A}=-\mathrm{tr}_{A}(\rho_{A}\ln\rho_{A})$, in which $\rho_A$ is the reduced density matrix of the subsystem $A$. In the context of the Anti-de Sitter/conformal field theory (AdS/CFT) correspondence~\cite{Maldacena:1997re,Gubser:1998bc,Witten:1998qj}, the entanglement entropy of a spatial subregion can be identified as the area of a codimension-two bulk minimal surface, which is termed as the Ryu--Takayanagi (RT) formula of the Holographic entanglement entropy (HEE)~\cite{Ryu:2006bv}. The HEE was later extended to a covariant form~\cite{Hubeny:2007xt} and derived from the gravitational entropy~\cite{Lewkowycz:2013nqa,Casini:2011kv,Dong:2016hjy}. The dual geometric interpretation of entanglement has provided important tools and insights for studying bulk reconstruction and the emergence of spacetime in gauge/gravity duality
\cite{Swingle:2009bg,Pastawski:2015qua,Dong:2016eik,Lin:2020ufd,Liang:2025vmx}. In particular, it was shown that the dynamics of entanglement entropy of the boundary CFT can be expressed in a first law like relation, called the entanglement first law~\cite{Bhattacharya_2013,Swingle:2014uza,Nozaki:2013vta,He:2014lfa,deBoer:2015kda,deBoer:2016pqk,Haehl:2017sot,Guo:2013aca}.
Furthermore, the entanglement first law can be reformulated in terms of the variations of the relative entropy and modular Hamiltonian~\cite{Blanco:2013joa}, which was shown to be equivalent to linearized Einstein equations in asymptotically AdS spacetime~\cite{Bhattacharya:2013bna,Lashkari:2013koa,Faulkner:2013ica}. This relation builds on the older connection between gravity and thermodynamics~\cite{Bekenstein:1973ur,Bardeen:1973gs,Hawking:1975vcx,Jacobson:1995ab} and has also been extended beyond linear order perturbation in Einstein gravity~\cite{Faulkner:2017tkh,Oh:2017pkr}.

Entanglement entropy is not the only measure to characterize entanglement, there are many other measures such as the R\'{e}nyi entropy~\cite{Hung:2011nu,Dong:2016fnf}, entanglement of purification~\cite{Takayanagi:2017knl,Lin:2020yzf}, entanglement contour~\cite{Wen:2018whg,Lin:2022aqf} and hyperfine structure of entanglement~\cite{Mo:2023kym}, et al and their holographic descriptions also been proposed. Apart from the well understood entanglement between spatial regions, entanglement measures have also been extended to observables associated with non-spacelike or transition-matrix data, called the pseudo entropy, which was introduced as the von Neumann entropy of a reduced transition matrix, and gives
a natural complex generalization of entanglement entropy~\cite{Nakata:2020luh,Mollabashi:2020yie,Mollabashi:2021xsd}. Timelike
entanglement entropy (TEE), where the boundary interval or subregion extends
along a timelike direction, can be defined in field theory through analytic
continuation and is closely tied to pseudo entropy~\cite{Doi:2022iyj,Doi:2023zaf}. In the dS/CFT correspondence and related analytic continuations, this connection has been developed through extremal surfaces and replica-like
constructions~\cite{Doi:2022iyj,Narayan:2022afv,Narayan:2023zen,Nanda:2025tid}.

On the holographic side, several complementary lines of development have
emerged. The first treats holographic timelike entanglement entropy (HTEE) as
an analytic continuation of the ordinary HEE or as a double Wick rotated RT
problem~\cite{Li:2022tsv,Doi:2022iyj,Doi:2023zaf,He:2023ubi}. This viewpoint
has been applied to the dS$_3$/CFT$_2$ duality, $T\bar{T}$-deformed theories, Rindler
methods, relations between timelike and spacelike entanglement, and the imaginary
part of TEE~\cite{Jiang:2023loq,Jiang:2023ffu,Guo:2024lrr,Xu:2024yvf,Maulik:2026vmj}. It has also been used to explore black hole singularities, causally connected subregions,
top-down backgrounds, interpolations between spacelike and timelike
entanglement, and higher dimensional setups~\cite{Anegawa:2024kdj,Gong:2025pnu,Nunez:2025gxq,Nunez:2025ppd,Nunez:2025puk,Jiang:2025pen}.
The second line emphasizes that the relevant saddles are, in general, complex
codimension-two extremal surfaces. This complex-surface interpretation gives
a geometric prescription for temporal entanglement and sharpens the relation
between timelike observables and ordinary holographic entanglement~\cite{Heller:2024whi,Heller:2025kvp}. While the third line studies pseudo entropy directly in holographic settings, including the real-time AdS/CFT correspondence, entanglement phase transitions in holographic pseudo entropy, thermal pseudo entropy, and their CFT derivations~\cite{Chen:2023holographicPseudo,Kanda:2023jyi,Caputa:2024gve,Kanda:2026pse}.
Related developments also connect timelike modular data to bulk
reconstruction in (A)dS backgrounds~\cite{Das:2023timelikeReconstruction}. More recently, the entanglement first law has been extended to the TEE and its equivalence between linearized Einstein equations in asymptotically AdS$_{d+1}$ (for $d\geq 2$) spacetime has also been proven~\cite{Li:2025tud}, and the entanglement first law of pseudo entropy in dS$_3$ spacetime has been studied in~\cite{Fujiki:2025rtx}.

Despite these progresses, the higher curvature and first-law aspects of HTEE are
still much less developed. Higher curvature interactions are expected in
effective descriptions of quantum gravity, and they will change the black hole area entropy as well as the holographic entropy functional: the area functional is replaced by generalized entropy functionals, with possible Wald-like and extrinsic-curvature contributions~\cite{Wald:1993nt,Jacobson:1993vj}. As in \cite{Sun:2008uf}, the author first discussed the modification of HEE due to the gravitational Chern-Simons term. In general, Wald's entropy does not give the correct HEE functional~\cite{Hung:2011xb}. For Lovelock gravity, however, the Jacobson-Myers (JM) entropy functional~\cite{Jacobson:1993xs} depends only on the intrinsic curvature of the bulk surface and passes nontrivial
anomaly checks~\cite{Hung:2011xb,deBoer:2011wk}. Recent work on HTEE in
higher curvature gravity, in particular, \cite{Zhao:2025zgm} found universal correction patterns for cubic Lovelock theories and showed that excitations contribute only to the real part in the setups considered there. Hyperbolic subsystems, which are the natural arena for a modular-Hamiltonian first law, require a separate treatment because excited-state perturbations spoil the diagonal structure used in simpler strip geometries. This motivates the present work: we ask whether the timelike entanglement first law survives in Lovelock gravity, and whether the same
coupling-dependent factor that appears in the linearized Lovelock field equations
also controls the variations of the TEE and the modular Hamiltonian.

The point of the analysis is not to rediscover the known universality of the
JM functional in ordinary Lovelock HEE. Rather, we isolate the corresponding
statement in the timelike entanglement first law setting, where the surface is obtained by double Wick rotation and the modular Hamiltonian is fixed by the hyperbolic subregion. We first verify the mechanism in cubic Lovelock gravity for a low-energy thermal excitation in the dual CFT. We then show, for arbitrary Lovelock order, that normalizable Fefferman-Graham (FG) perturbations about AdS spacetime give the same overall factor in $\Delta S$ and in $\Delta\langle H\rangle$, while the generalized boundary term does not contribute in the conformal limit. Our results provide a restricted but rigorous generalization of the equivalence between the timelike entanglement first law and the linearized gravitational field equations, extending it from Einstein gravity to the broader Lovelock class.

Let us also emphasize the domain of validity. Throughout the paper we work
perturbatively around the AdS vacuum, keep the boundary geometry fixed and
conformally flat, and consider hyperbolic timelike subregions for which the
modular Hamiltonian is local. The result should therefore be viewed as a
statement about this universality class, rather than as a general theorem for
arbitrary time-dependent states, non-hyperbolic regions or generic complex
extremal saddles.

The paper is organized as follows. In section~\ref{sec2} we briefly review Lovelock
gravity and its relevant entropy functionals, as well as the timelike entanglement first law in Einstein gravity. In section~\ref{sec3} we test the first law for low-energy thermal excitations in cubic Lovelock gravity. In section~\ref{sec4} we derive the Lovelock field equations linearized about a maximally symmetric AdS spacetime background. Section~\ref{sec5} gives the general analysis for linear perturbations in
FG gauge and for Lovelock theories of arbitrary order. Conclusions and discussions are drawn in section~\ref{sec6}. Besides, some technical details are collected in the
appendix.

\textbf{Notations:} Uppercase Latin indices such as $A,B,C$ denote bulk
spacetime indices. Greek indices such as $\mu,\nu,\rho$ denote boundary
indices. Lowercase Latin indices such as $i,j,k$ denote coordinates on the
extremal surface, and $a,b,c$ denote angular coordinates on the 
entangling surface. The metric $g_{AB}$ denotes the bulk metric including the
conformal factor, while $h_{\mu\nu}$ denotes the corresponding boundary
metric. On the extremal surface, $\gamma_{ij}$ denotes the induced metric including
the conformal factor, and $\sigma_{ab}$ denotes the metric on the spherical
entangling surface on the AdS boundary. A tilde, as in $\tilde{g}_{AB}$, denotes the corresponding metric without the conformal factor. Hatted indices $\hat m,\hat n$ label the two normal directions to a codimension-two surface, and
$\eta_{\hat m\hat n}$ is the metric on this normal bundle.

\section{A few preliminaries}\label{sec2}
The primary objective of this work is to examine how higher curvature
interactions in the bulk gravitational theory affect the timelike
entanglement first law. We work in Lovelock gravity
\cite{Lovelock:1970zsf,Lovelock:1971yv}, a tractable higher curvature
model in which explicit calculations can be performed without introducing
higher derivative equations of motion. We begin by reviewing the aspects
of Lovelock gravity needed below. We then recall two entropy functionals
which are relevant for higher curvature holography: Wald's entropy
formula \cite{Wald:1993nt,Iyer:1994ys,Jacobson:1993vj}, valid for
general covariant gravitational actions, and the Jacobson-Myers functional
\cite{Jacobson:1993xs}, which is tailored to Lovelock gravity. Finally,
we review the timelike entanglement first law in the double Wick rotation
formalism of \cite{Li:2025tud}.

\subsection{Lovelock gravity}\label{lovelock gravity}
Lovelock gravity \cite{Lovelock:1970zsf,Lovelock:1971yv} is a
higher dimensional theory in which the higher curvature interactions are
Euler densities of even dimensional manifolds. The general Lovelock action
in $d+1$ dimensions can be written as
\begin{equation}
I=\frac{1}{16\pi G_{d+1}}\int d^{d+1}x\sqrt{-g}
\sum_{m\ge 0}^{\lfloor\frac{d+1}{2}\rfloor}
c_{m}L^{2m-2}\mathcal{L}_{m}(R),\label{action}
\end{equation}
where $G_{d+1}$ is the $(d+1)$-dimensional Newtonian constant, $[(d+1)/2]$ denotes the integer part of $(d+1)/2$
and $c_{m}$ are dimensionless coupling constants for the higher curvature
terms $\mathcal{L}_{m}(R)$. For instance, $c_0=d(d-1)$ and $c_1=1$ give the negative cosmological constant and the Einstein-Hilbert terms, respectively. These higher derivative interactions are
defined as
\begin{align}
\mathcal{L}_{m}(R)\equiv\frac{1}{2^{m}}\delta_{B_{1}B_{2}\ldots B_{2m-1}B_{2m}}^{A_{1}A_{2}\ldots A_{2m-1}A_{2m}}R_{A_{1}A_{2}}^{ \phantom{B_1B_2} B_{1}B_{2}}\cdots R_{A_{2m-1}A_{2m}}^{\phantom{B_{2m-1}B_{2m}} B_{2m-1}B_{2m}},   
\end{align}
which is proportional to the Euler density on a $2m$-dimensional
manifold. Here, the symbol $\delta_{B_{1}B_{2}\ldots B_{2m}}^{A_{1}A_{2}\ldots A_{2m}}$
is used to denote the totally antisymmetric product of $2m$ Kronecker
delta functions. 

By construction, in $(d+1)$ dimensions all Lovelock $\mathcal{L}_{m}$ terms
with $m>(d+1)/2$ vanish, so the explicit restriction on the sum in
\eqref{action} is not essential. For $m=(d+1)/2$, $\mathcal{L}_{m}$ is topological
and does not contribute to the gravitational equations of motion.

The original motivation to construct this action \eqref{action} was that the
resulting gravitational equations of motion are only second order in derivatives
\cite{Lovelock:1970zsf,Lovelock:1971yv}. In anticipation of applications
to the AdS/CFT correspondence, a negative cosmological constant is
explicitly included in the action. The theory then admits a pure $\textrm{AdS}_{d+1}$ solution, 

\begin{equation}
ds^{2}=\frac{\tilde{L}^{2}}{z^{2}}\big(-dt^{2}+dz^{2}+dx_{d-1}^{2}+d\mathbf{x}^{2}\big),\quad\mathbf{x}\in\mathbb{R}^{d-2},\label{pure AdS-1-1}
\end{equation}
where $\tilde{L}$ is the effective curvature radius of AdS spacetime in Lovelock gravity, defined by $\tilde{L}^{2}=L^{2}/f_{\infty}$, and $f_{\infty}$ is a root of
\begin{align}
1=f_{\infty}-\sum_{m=2}^{\lfloor d/2\rfloor}\lambda_{m}(f_{\infty})^{m},    
\end{align}
and the coefficients $\lambda_{m}$ are defined as
\begin{align}
\lambda_{m}=(-)^{m}\frac{(d-2)!}{(d-2m)!}c_{m}.    
\end{align}

In a regime where the $\lambda_{m}$ are not large, the relevant solution
is the smallest positive root, and it is continuously connected to
the single root $f_{\infty}=1$ that remains in the Einstein gravity
limit, i.e., $\lambda_{m}\rightarrow0$. To ensure a smooth connection
with the Einstein gravity limit while capturing higher derivative
gravitational corrections to timelike entanglement entropy, the discussion
is confined to this small coupling regime and with this particular
root in the following.

\subsection{Holographic timelike entanglement entropy in higher curvature gravity }
Consider a black hole in a $(d+1)$-dimensional spacetime $M$ governed by
Lovelock gravity. There are two closely related entropy functionals that will
be relevant below. The first is the Wald entropy, obtained from the Lagrangian
variation of a generally covariant theory
\cite{Wald:1990mme,Wald:1993nt,Wald:1999vt,KAY199149,Lee:1990nz,Iyer:1994ys}:
\begin{align}
    S=2\pi\int_{\Sigma} \textbf{X}^{AB}\epsilon_{AB},\label{Wald}
\end{align}
where $\Sigma$ is the spatial section of the horizon (the bifurcation surface), $(\textbf{X}^{AB})_{C_3\dots C_{d+1}}=\frac{\partial{\rl}}{\partial R_{ABCD}}\boldsymbol{\epsilon}_{CDC_3\dots C_{d+1}}$, with $\boldsymbol{\epsilon}_{CDC_3\dots C_{d+1}}$ denoting the volume form of $M$, and $\epsilon_{AB}$ represents the binormal volume form of the tangent space perpendicular to $\Sigma$. 
Written explicitly for Lovelock theory, the Wald entropy is given by
\begin{align}
    S_{\rm W}&=\frac{1}{4G_{d+1}}\int _{\Sigma} d^{d-1}x\sqrt{\gamma}\Big[1+\sum_{m=2}^{[\frac{d+1}{2}]}m c_m L^{2m-2}\rl_{m-1}(R^{\parallel}[g])\Big],
\end{align}
where $\gamma$ is the determinant of the induced metric on $\Sigma$, and
$R^{\parallel}[g]$ is the bulk Riemann tensor projected onto $\Sigma$.
From the Gauss-Codazzi equations, the projected curvature can be related to
the intrinsic curvature $\r_{ijkl}[\gamma]$ of $\Sigma$ as
\begin{align}
    R^{\parallel}_{ijkl}[g]=\r_{ijkl}[\gamma]-\sum_{\hat m,\hat n=1}^2\eta_{\hat m\hat n}\Big(\mathfrak{K}^{\hat m}_{ik}\mathfrak{K}^{\hat n}_{jl}-\mathfrak{K}^{\hat m}_{il}\mathfrak{K}^{\hat n}_{jk}\Big),
\end{align}
where $\mathfrak{K}_{ij}^{\hat m}$ denotes the extrinsic curvature associated
with the normal directions of $\Sigma$.

The second is the JM entropy functional, which follows from the
Hamiltonian analysis of Lovelock gravity \cite{Jacobson:1993xs}:
\begin{align}
    S_{\rm JM}=\frac{1}{4G_{d+1}}\int _{\Sigma} d^{d-1}x\sqrt{\gamma}\Big[1+\sum_{m=2}^{[\frac{d+1}{2}]}m c_m L^{2m-2}\rl_{m-1}(\r[\gamma])\Big].\label{JM}
\end{align}
The difference between these two formulas is controlled by the extrinsic
curvature and therefore vanishes for stationary black hole horizons, where
$\mathfrak{K}_{ij}^{\hat m}=0$. For a generic holographic entangling surface,
however, the full second fundamental form need not vanish, and one must decide
which functional gives the correct entropy by an independent check.

The conformal trace anomaly provides such a check \cite{Hung:2011xb}. In a
pure AdS background, extremizing Wald's entropy functional reduces to
extremizing an area functional with an overall coefficient proportional to
$a_d^*$, the A-type anomaly coefficient in even boundary dimensions. The
universal logarithmic term obtained from Wald's functional is therefore
proportional only to the A-type anomaly. Field theoretic calculations for a
generic entangling surface instead give a geometry dependent combination of
A-type and B-type anomaly coefficients. This mismatch rules out Wald's entropy
as the general HEE functional in higher curvature gravities.

For Lovelock gravity, the same analysis shows that $S_{\rm JM}$ reproduces the
expected A-type and B-type anomaly data for boundary CFTs in $d=4$ and $d=6$
and for a variety of boundary geometries and entangling surfaces
\cite{Hung:2011xb}. This provides strong evidence that the JM functional,
rather than Wald's entropy functional, is the appropriate HEE functional for
Lovelock gravity.

For HTEE the situation is more subtle, because the relevant extremal surfaces
are obtained by analytic continuation and are generically complex
\cite{Doi:2023zaf,Heller:2024whi,Heller:2025kvp}. This suggests that the RT
prescription, originally formulated for real spacelike surfaces, should be
continued to the complex saddles appropriate to timelike anchoring. In
higher curvature gravity this continuation should act not only on the
embedding but also on the entropy functional. Ref.~\cite{Zhao:2025zgm}
implemented this idea for the JM functional in Lovelock gravity and found
universal correction patterns; in particular, the excited state contribution
appears only in the real part in the setups studied there. The hyperbolic
subsystems considered in the present paper require a different treatment,
because excited-state terms break the diagonal structure of the induced metric
on the extremal surface. The double Wick rotation method of \cite{Li:2025tud}
provides an effective way to compute HTEE for excited hyperbolic subsystems,
and in the vacuum it reproduces the result obtained from complexified extremal
surfaces. These observations motivate using the JM functional \eqref{JM} in
the double Wick rotated geometry to analyze HTEE in Lovelock gravity.

In what follows we adopt the following working prescription. Starting from the
Lovelock HEE functional \eqref{JM}, we perform the same double Wick rotation
on the background geometry and on the anchoring data that defines the timelike
subregion. The resulting functional is then extremized in the continued
geometry, and its first variation is compared with the CFT modular Hamiltonian
for the hyperbolic region. Equivalently, this prescription selects the
Einstein-continuous branch of complex extremal surfaces in a neighborhood of
the AdS vacuum. We do not attempt to derive this rule from a Lorentzian
replica construction for general higher curvature gravity. The results below
should therefore be read as consequences of this precise timelike continuation
prescription, within the perturbative regime in which the double Wick rotated
description is under control.

\subsection{Timelike entanglement first law}

The relative entropy between two states in the same Hilbert space provides a
fundamental measure of their distinguishability. Given two density matrices
$\rho_1$ and $\rho_0$, the relative entropy $S(\rho_1|\rho_0)$ is defined as
\begin{align}
S(\rho_1|\rho_0)={\rm tr}(\rho_1\ln\rho_1)-{\rm tr}(\rho_1\ln\rho_0).    
\end{align}
Relative entropy is positive and monotonic under inclusion. For two
infinitesimally different states, its first-order variation vanishes and yields
the entanglement first law
\begin{align}\label{1stlaw}
    \Delta S= \Delta\langle H\rangle,
\end{align}
where $\Delta S$ is the first-order variation of the entanglement entropy, and
$H$ is the modular Hamiltonian associated with the subsystem, defined by
\begin{equation}
    \rho=\frac{e^{-H}}{{\rm tr}e^{-H}},
\end{equation}
and $\langle H\rangle={\rm tr}(\rho H)$ is its expectation value. The general relation \eqref{1stlaw} is valid for arbitrary perturbations of an arbitrary state on any spatial region $A$.
Early holographic studies focused on relative entropy between the vacuum and
excited states for spherical regions on a spatial boundary. In this case, the
domain of dependence of the ball-shaped region can be mapped by a conformal
transformation to hyperbolic spacetime. As shown in
\cite{Casini:2011kv}, such a transformation maps the vacuum density matrix for the region $A$ to the
thermal density matrix for the hyperbolic space theory. Mapping back to the
ball-shaped region of Minkowski space, it follows \cite{Casini:2011kv} that the modular Hamiltonian can be written as
\begin{align}
      H_{\mathcal{D}}&=2\pi \int_{r\le R}d^{d-1}x \frac{\left(R^2-r^2\right)}{2R}T^{00}(\mathbf{x}),
\end{align}
In any CFT with a semiclassical holographic dual, this first law has a
gravitational interpretation as a constraint on the bulk spacetime dual to the
CFT state. For small perturbations around the CFT vacuum, the set of such
constraints for all ball-shaped spatial regions is equivalent to the requirement
that the dual geometry satisfy the gravitational equations of motion linearized
about the pure AdS spacetime.

Recent work \cite{Li:2025tud} generalized this spacelike entanglement
statement to the timelike case. By implementing a double Wick rotation, the
exact modular Hamiltonian for the hyperbolic subsystem in the boundary CFT is
obtained as \cite{Li:2025tud}
\begin{align}   
H_{\mathcal{D}'} & =2\pi\int_{\mathcal{D}'}d\tau d^{d-2}\mathbf{x}\,
\frac{\left(T_0^{2}-\tau^{2}-\mathbf{x}^{2}\right)}{2T_0}
T^{(d-1)(d-1)},\label{delta H}
\end{align} 
where $\mathcal{D}'=\{(\tau,\mathbf{x})\,|\,\tau^2+\mathbf{x}^2\le T_0^2\}$
and $T_0$ sets the size of the timelike region. This makes it possible to formulate the timelike version of \eqref{1stlaw}
from the variation of relative entropy. In the double Wick rotated setting,
ref.~\cite{Li:2025tud} also proved the equivalence between the linearized
Einstein equations and the timelike entanglement first law, revealing a direct
link between timelike entanglement dynamics and bulk gravity.

\section{Simple example testing the timelike entanglement first law}\label{sec3}
As described above, our strategy is to test the equality \eqref{1stlaw} in a
holographic setting with higher curvature interactions for a hyperbolic
entangling surface, for which the modular Hamiltonian \eqref{delta H} is known.
In the double Wick rotated geometry, the JM prescription allows us to compute
the timelike entanglement entropy in Lovelock gravity and hence $\Delta S$.
The same setup also allows us to evaluate $\Delta\langle H\rangle$ from the
boundary stress tensor. In this section we focus on cubic Lovelock gravity,
namely on the truncation of \eqref{action} to $m_{\textrm{max}}=3$, where all
quantities can be computed explicitly. The reference state is the CFT vacuum,
while the perturbed state is dual to a black hole excitation of the AdS
background. This tractable example prepares the more general analysis of
section \ref{sec5}.
\subsection{Low-energy thermal excitation in cubic Lovelock gravity}
We now begin with the case of the cubic Lovelock gravity in $(d+1)$ dimensions.
For a $(d+1)$-dimensional bulk, the curvature-squared and curvature-cubed
interactions contribute to the Lovelock action \eqref{action}, yielding
\begin{align}
I&=\frac{1}{16\pi G_{d+1}}\int d^{d+1}x\sqrt{-g}
\left[\frac{d(d-1)}{L^{2}}+R
+\frac{L^{2}\lambda}{(d-2)(d-3)}\mathcal{L}_{4}(R)\right. \notag\\
&\hspace{6.2cm}\left.
{}-\frac{L^{4}\mu}{(d-2)(d-3)(d-4)(d-5)}
\mathcal{L}_{6}(R)\right].
\end{align}

Comparing with the notation of section \ref{lovelock gravity}, we have $\lambda=\lambda_{2}=(d-2)(d-3)c_{2}$
and $\mu=\lambda_{3}=-(d-2)(d-3)(d-4)(d-5)c_{3}$.

Cubic Lovelock gravity admits the pure AdS solution \eqref{pure AdS-1-1}, where $f_{\infty}=L^{2}/\tilde{L}^{2}$ is the smallest positive root of
\begin{align}
1=f_{\infty}-\lambda f_{\infty}^{2}-\mu f_{\infty}^{3}.   
\end{align}

For cubic Lovelock gravity in $(d+1)$ dimensions, the HEE has been discussed in \cite{Hung:2011xb,deBoer:2011wk}, which
can be expressed as
\begin{align}
S&=\frac{1}{4G_{d+1}}\int_{\mathcal{M}}d^{d-1}x\sqrt{\gamma}
\left[1+\frac{2L^{2}\lambda}{(d-2)(d-3)}\mathcal{R}\right. \notag\\
&\hspace{4.6cm}\left.
{}-\frac{3L^{4}\mu}{(d-2)(d-3)(d-4)(d-5)}
\mathcal{L}_{4}(\mathcal{R})\right]+S_{\textrm{surf}},\label{HEE-1}
\end{align}
where $\gamma$ is the determinant of the induced metric $\gamma_{ij}$ on the bulk surface
$\mathcal{M}$, and $\mathcal{R}_{ijkl}$ is the Riemann
tensor of $\mathcal{M}$. The surface term is added to ensure a well-defined
variational principle in extremizing the functional, which is~\cite{Myers:1987yn,Myers:2013lva}
\begin{align}
S_{\textrm{surf}}
&=\frac{1}{4G_{d+1}}\int_{\partial\mathcal{M}}d^{d-2}x\sqrt{\sigma}
\left[\frac{4\lambda L^{2}}{(d-2)(d-3)}\mathcal{K}\right.\label{surface term-1}\notag\\
&\quad\left.
{}-\frac{3L^{4}\mu}{(d-2)(d-3)(d-4)(d-5)}
\left(4R^{\partial}\mathcal{K}
-8R^{\partial\,ab}\mathcal{K}_{ab}
-\frac{4}{3}\mathcal{K}^{3}\right.\right. \notag\\
&\quad\left.\left.
+4\mathcal{K}\mathcal{K}_{ab}\mathcal{K}^{ab}
-\frac{8}{3}\mathcal{K}_{ab}\mathcal{K}^{bc}\mathcal{K}_{c}^{a}
\right)\right],
\end{align}
in which $\sigma$ is the determinant of the induced $\sigma_{ab}$ on the boundary of $\mathcal{M}$, i.e., $\partial\mathcal{M}$, $\mathcal{K}_{ab}$ is the extrinsic curvature of $\partial\mathcal{M}$ and $\mathcal{K}=\sigma^{ab}e^i_ae^j_b\nabla_i n_j=(\gamma^{ij}-n^in^j)\nabla_in_j=\gamma^{ij}\nabla_in_j$, while $R_{ab}^{\partial}$ and $R^{\partial}$ are the intrinsic Ricci tensor and Ricci scalar of
$\partial\mathcal{M}$, respectively.

At present, there is no general formula for HTEE in higher derivative gravity. Inspired by the proposals of \cite{Heller:2024whi,Zhao:2025zgm,Li:2025tud}, timelike entanglement entropy can be associated either with a complex extremal surface or with an extremal surface obtained by double Wick rotation. These observations provide the basis for extending the RT prescription, originally proposed for real spacelike extremal surfaces, to the extremal surfaces appearing in eq.~\eqref{HEE-1}, and hence for giving a holographic interpretation of TEE in cubic Lovelock gravity.

Our purpose is to compute the variation of TEE for a low-energy excited CFT state from the holographic point of view, in which the low-energy excited CFT state is dual to a bulk asymptotically AdS spacetime, while the pure AdS solution \eqref{pure AdS-1-1} is the ground state of the AdS Lovelock gravity. An excitation of pure AdS in cubic Lovelock gravity can be described by
\begin{align}
ds^{2}=\frac{\tilde{L}^{2}}{z^{2}}\left[\frac{1}{f(z)}dz^{2}-f(z)dt^{2}+dx_{d-1}^2+d\mathbf{x}^{2}\right],\mathbf{x}\in\mathbb{R}^{d-2}\label{excited state}
\end{align}

with $f(z)\approx 1-m_zz^{d}$, where $m_z$ characterizes the near boundary
deviation from the pure AdS metric. We identify this perturbation as a low-energy thermal excitation and, unless stated otherwise, all higher curvature
gravitational corrections in this section refer to the TEE in such excited states. It is challenging to obtain an exact expression for the TEE in
a black hole background within Lovelock gravity. To make progress,
the excitation parameter $m_z$ is treated as a small perturbative parameter
and the method of \cite{Guo:2013aca,Li:2025tud} can be adopted. Since
the variation of the shape of the bulk surface does not contribute
to the variation of the functional \eqref{HEE-1} up to the order $O(m_z)$, the linear
correction to \eqref{HEE-1} after the excitation is turned on is entirely
determined by the metric perturbation evaluated on the same extremal
surface when $m_z=0$.
\subsection{\texorpdfstring{$\Delta S=\Delta \langle H \rangle$}{Delta S = Delta <H>} for low-energy thermal excitation}
The hyperbolic subsystem is defined as
$A=\left\{ \left.(t,\mathbf{x},x_{d-1})\right|-t^{2}+\mathbf{x}^{2}<-T_{0}^{2},x_{d-1}=0\right\} $
in a constant space slice of the CFT living on the boundary. We compute
the TEE by applying the double Wick rotation
$t\rightarrow-i\tau$, $x_{d-1}\rightarrow ix_{d-1}$ following the notations
of \cite{Li:2025tud}. The induced metric on the Wick rotated bulk
surface is
\begin{equation}
ds^{2}=\frac{\tilde{L}^{2}}{z^{2}}\left[(1+m_zz^{d})dz^{2}+(1-m_zz^{d})d\tau^{2}+d\textbf{x}^{2}\right].\label{induced metric-1}
\end{equation}
Introducing a radial coordinate $\xi^{2}=\tau^{2}+\textbf{x}^{2}$
and defining $\tau=\xi\cos\theta,|\mathbf{x}|=\xi\sin\theta$,
the surface that extremizes the functional \eqref{HEE-1} when the perturbation
parameter $m_z$ is turned off can be parameterized as \cite{Hung:2011xb,Guo:2013aca}
\begin{equation}
\xi(u)=T_{0}\cos(u/T_{0}),\quad z(u)=T_{0}\sin(u/T_{0}),\quad\epsilon\le u\le\frac{\pi}{2}T_{0}.
\end{equation}
The induced metric \eqref{induced metric-1} with $m_z=0$ becomes
\begin{equation}
ds^{2}=\frac{\tilde{L}^{2}}{\sin^{2}w}\left[dw^{2}+\cos^{2}w\left(d\theta^{2}+\sin^{2}\theta\,\tilde{\chi}_{a'b'}d\Omega^{d-3}\right)\right],a',b'=1,\dots,d-3\label{maximally symmetric}
\end{equation}
where $w=u/T_{0},$ and $\tilde{\chi}_{a'b'}$ is the metric
of the $(d-3)$-dimensional unit sphere. It is straightforward to show that \eqref{maximally symmetric}
is a $(d-1)$-dimensional maximally symmetric spacetime with
\begin{align}
\bar{\mathcal{R}} & =-\frac{(d-1)(d-2)}{\tilde{L}^{2}},\label{bar R}\\
\bar{\mathcal{L}_{4}} & =\frac{(d-1)(d-2)(d-3)(d-4)}{\tilde{L}^{4}},\label{bar L4}
\end{align}    
where the bars mean that the corresponding quantities are evaluated on the background metric.

When the excitation is turned on, the induced metric \eqref{induced metric-1}
then becomes
\begin{equation}
\begin{aligned}\gamma_{uu} & =\frac{\tilde{L}^{2}}{T_{0}^{2}\sin^{2}w}\left[1+m_zT_{0}^{d}\sin^{d}w(\cos^{2}w-\sin^{2}w\cos^{2}\theta)\right],\\
\gamma_{u\theta} & =-m_zT_{0}^{d-1}\tilde{L}^{2}\sin^{d-1}w\cos w\cos\theta\sin\theta,\\
\gamma_{\theta\theta} & =\frac{\tilde{L}^{2}\cos^{2}w}{\sin^{2}w}\left[1-m_zT_{0}^{d}\sin^{d}w\sin^{2}\theta\right],\\
\gamma_{a'b'} & =\frac{\tilde{L}^{2}\cos^{2}w}{\sin^{2}w}\sin^{2}\theta\,\tilde{\chi}_{a'b'}(\Omega),
\end{aligned}
\end{equation}
and the determinant of the induced metric on the hypersurface is
\begin{align}
\sqrt{\gamma}
&=\frac{\tilde{L}^{d-1}\cos^{d-2}w\sin^{d-3}\theta}
{T_{0}\sin^{d-1}w}\sqrt{\tilde{\chi}}
\Big[1+\frac{1}{2}m_zT_{0}^{d}\sin^{d}w
 \notag\\
&\hspace{4.1cm}
\times(\cos^{2}w-\sin^{2}w\cos^{2}\theta-\sin^{2}\theta)
+O(m_z^{2})\Big].\label{square h}
\end{align}
Denoting $\gamma_{ij}=\bar{\gamma}_{ij}+\delta \gamma_{ij}$, where the
background $\bar{\gamma}_{ij}$ is in eq.~\eqref{maximally symmetric}, then
$\delta \gamma_{ij}$ is given by
\begin{equation}
\begin{aligned}\delta \gamma_{uu} & =\frac{\tilde{L}^{2}}{T_{0}^{2}\sin^{2}w}\left[m_zT_{0}^{d}\sin^{d}w(\cos^{2}w-\sin^{2}w\cos^{2}\theta)\right],\\
\delta \gamma_{u\theta} & =-\tilde{L}^{2}m_zT_{0}^{d-1}\sin^{d-1}w\cos w\cos\theta\sin\theta,\\
\delta \gamma_{\theta\theta} & =-\frac{\tilde{L}^{2}\cos^{2}w}{\sin^{2}w}m_zT_{0}^{d}\sin^{d}w\sin^{2}\theta,\\
\delta \gamma_{a'b'} & =0.
\end{aligned}
\label{perturbation}
\end{equation}

Then the linear correction of the Ricci scalar and the Gauss-Bonnet term
for the intrinsic geometry of the bulk surface $\mathcal{M}$ is given
by
\begin{align}
\delta\mathcal{R} & =\frac{d-2}{\tilde{L}^{2}}\delta \gamma+\nabla_{i}\nabla_{j}\delta \gamma^{ij}-\nabla^{2}\delta\gamma,\label{linear R} \\
\delta\mathcal{L}_{4} & =\frac{-2(d-3)(d-4)}{\tilde{L}^{2}}\delta\mathcal{R}.\label{delta L4}
\end{align}
where $\delta \gamma=\bar{\gamma}^{ij}\delta \gamma_{ij},$ $\delta \gamma^{ij}=\bar{\gamma}^{ik}\bar{\gamma}^{jl}\delta \gamma_{kl}$,
and the covariant derivatives $\nabla$ is compatible with the background
metric $\bar{\gamma}$. Explicit evaluating \eqref{linear R} with \eqref{maximally symmetric}
and \eqref{perturbation} gives 
\begin{equation}
\delta\mathcal{R}=\frac{m_zT_{0}^{d}\sin^{d}w}{\tilde{L}^{2}}\left[(2d+1)\cos^{2}w-(d^{2}+d-1)\cos^{2}w\cos^{2}\theta-d+2\right].\label{delta R-1}
\end{equation}

Now we can turn to consider the surface terms in \eqref{surface term-1}.
The normal outward unit vector to the boundary surface defined by
$u=u_{\epsilon}$, or equivalently  $\xi(u_{\epsilon})=T_{0},z(u_{\epsilon})=\epsilon$,
is given by
\begin{equation}
n_{i}=-\frac{1}{\sqrt{\gamma^{uu}}}\delta_{iu},
\end{equation}
where $i$ runs over $u,\theta$ and a $(d-3)$-dimensional sphere.
Thus all the components of the extrinsic curvature $\left.\mathcal{K}_{ij}=\nabla_{i}n_{j}\right|_{u=u_{\epsilon}}$ can be obtained as follows explicitly\footnote{To avoid ambiguity, the following expressions are all evaluated at
the boundary $u=u_{\epsilon}$. For brevity, we will omit the notation$\ensuremath{\left.\right|_{u=u_{\epsilon}}}$
hereafter.} 
\begin{equation}
\begin{aligned}
\mathcal{K}_{uu} & =-\frac{1}{2}\sqrt{\frac{\gamma_{\theta\theta}}{H}}\,\partial_{u}\!\left(\frac{H}{\gamma_{\theta\theta}}\right)+
\frac{1}{2}\Bigl(\gamma^{uu}\partial_{u}\gamma_{uu}+2\gamma^{u\theta}\partial_{u}\gamma_{u\theta}-\gamma^{u\theta}\partial_{\theta}\gamma_{uu}\Bigr)\sqrt{\frac{H}{\gamma_{\theta\theta}}},\\
\mathcal{K}_{u\theta} & =\frac{1}{2}\Bigl(\gamma^{uu}\partial_{\theta}\gamma_{uu}+\gamma^{u\theta}\partial_{u}\gamma_{\theta\theta}\Bigr)\sqrt{\frac{H}{\gamma_{\theta\theta}}},\quad \mathcal{K}_{\theta u}=-\frac{1}{2}\sqrt{\frac{\gamma_{\theta\theta}}{H}}\,\partial_{\theta}\!\left(\frac{H}{\gamma_{\theta\theta}}\right)+\mathcal{K}_{u\theta},\\
\mathcal{K}_{\theta\theta} & =\frac{1}{2}\Bigl(2\gamma^{uu}\partial_{\theta}\gamma_{u\theta}-\gamma^{uu}\partial_{u}\gamma_{\theta\theta}+\gamma^{u\theta}\partial_{\theta}\gamma_{\theta\theta}\Bigr)\frac{1}{\sqrt{\gamma^{uu}}},\\
\mathcal{K}_{a'b'} & =-\frac{1}{2}\Bigl(\sqrt{\gamma^{uu}}\,\partial_{u}E^{2}+\frac{\gamma^{u\theta}}{\sqrt{\gamma^{uu}}}\,\partial_{\theta}E^{2}\Bigr)\tilde{\chi}_{a'b'},\\
\mathcal{K}_{ua'} & =\mathcal{K}_{a'u}=\mathcal{K}_{\theta a'}=\mathcal{K}_{a'\theta}=0.
\end{aligned}    
\end{equation}
And the extrinsic curvature is
\begin{align}
\mathcal{K} & =\nabla_{i}n^{i}=\frac{1}{\sqrt{\gamma}}\partial_{i}\left(\sqrt{\gamma}n^{i}\right),\notag\\
 & =\frac{1}{\sqrt{H}E^{d-3}}\left[-\partial_{u}\left(E^{d-3}\frac{\sqrt{\gamma_{uu}}\gamma_{\theta\theta}}{\sqrt{H}}\right)
 +\partial_{\theta}\left(E^{d-3}\frac{\sqrt{\gamma_{uu}}\gamma_{u\theta}}{H}\right)\right],
\end{align}
where 
\begin{align}
H=\gamma_{uu}\gamma_{\theta\theta}-\gamma_{u\theta}^{2},\quad E^{2}=\frac{\tilde{L}^{2}\cos^{2}w}{\sin^{2}w}\sin^{2}\theta.    
\end{align}
Denote $\alpha=m_zT_{0}^{d}\sin^{d}w$, $\Theta^{2}=\frac{\tilde{L}^{2}\cos^{2}w}{\sin^{2}w}$.
Then the nonzero components of the intrinsic Ricci tensor and Ricci
scalar of the boundary $\partial\mathcal{M}$ can be calculated as
follows
\begin{align}
R_{\theta\theta}^{\partial} & =(d-3)\frac{(1-\alpha)}{1-\alpha\sin^{2}\theta},\\
R_{a'b'}^{\partial} & =\frac{(1-\alpha)}{(1-\alpha\sin^{2}\theta)^{2}}\Bigl((d-3)-\alpha(d-4)\sin^{2}\theta\Bigr)\sin^{2}\theta\tilde{\chi}_{a'b'},\\
R^{\partial} & =\frac{(d-3)(1-\alpha)}{\Theta^{2}(1-\alpha\sin^{2}\theta)^{2}}\Bigl[d-2-\alpha(d-4)\sin^{2}\theta\Bigr].
\end{align}
As a consistent check, when the excitation is turned off the boundary
reduces to a $(d-2)$-dimensional sphere with radius $\Theta$, i.e.,
$R^{\partial}=\frac{(d-3)(d-2)}{\Theta^{2}}$ when $\alpha=0$.

The surface terms in \eqref{surface term-1} are listed as
\begin{align}
R^{\partial\,ab}\mathcal{K}_{ab} & =\frac{(d-3)(1-\alpha)\mathcal{K}^{\theta\theta}}{(1-\alpha\sin^{2}\theta)}\notag\\
&+\frac{(1-\alpha)(d-3)\big(d-3-\alpha(d-4)\sin^{2}\theta\big)}{(1-\alpha\sin^{2}\theta)^{2}}\frac{\sin^{2}\theta\mathcal{K}_{S}}{E^{2}},\\
\mathcal{K}\mathcal{K}_{ab}\mathcal{K}^{ab} & =\mathcal{K}\left(\mathcal{K}_{\theta\theta}\mathcal{K}^{\theta\theta}+(d-3)\frac{\mathcal{K}_{S}^{2}}{E^{2}}\right),\\
\mathcal{K}_{ab}\mathcal{K}^{bc}\mathcal{K}_{c}^{a} & =\mathcal{K}_{\theta\theta}\mathcal{K}^{\theta\theta}\mathcal{K}_{\theta}^{\theta}+(d-3)\frac{\mathcal{K}_{S}^{3}}{E^{3}}.
\end{align}
where
\begin{align}
\mathcal{K}_{S}=-\frac{1}{2}\bigg(\sqrt{\gamma^{uu}}\,\partial_{u}E^{2}+\frac{\gamma^{u\theta}}{\sqrt{\gamma^{uu}}}\,\partial_{\theta}E^{2}\bigg). 
\end{align}
Substituting the explicit expression for the induced metric, we obtain
the specific first-order result as:
\begin{equation}
\begin{aligned}
\mathcal{K}  = & \frac{1}{\tilde{L}\cos w}\left[(d-2)-\frac{m_zT_{0}^{d}\sin^{d}w}{2}\left(d\cos^{2}\theta-2\right)\right],\\
R^{\partial}\mathcal{K} =&  \frac{(d-3)(d-2)\sin^{2}w}{\tilde{L}^{3}\cos^{3}w}\left[(d-2)-\frac{3m_zT_{0}^{d}\sin^{d}w}{2}\left(d\cos^{2}\theta-2\right)\right],\\
R^{\partial\,ab}\mathcal{K}_{ab}  = & \frac{(d-3)\sin^{2}w}{\tilde{L}^{3}\cos^{3}w}\left[(d-2)-\frac{3m_zT_{0}^{d}\sin^{d}w}{2}\left(d\cos^{2}\theta-2\right)\right],\\
\mathcal{K}\mathcal{K}_{ab}\mathcal{K}^{ab}  = & \frac{(d-2)}{\tilde{L}^{3}\cos^{3}w}\left[(d-2)-\frac{3m_zT_{0}^{d}\sin^{d}w}{2}\left(d\cos^{2}\theta-2\right)\right],\\
\mathcal{K}_{ab}\mathcal{K}^{bc}\mathcal{K}_{c}^{a}  = & \frac{1}{\tilde{L}^{3}\cos^{3}w}\left[(d-2)-\frac{3m_zT_{0}^{d}\sin^{d}w}{2}\left(d\cos^{2}\theta-2\right)\right],\\
\mathcal{K}^{3} = & \frac{(d-2)^{2}}{\tilde{L}^{3}\cos^{3}w}\left[(d-2)-\frac{3m_zT_{0}^{d}\sin^{d}w}{2}\left(d\cos^{2}\theta-2\right)\right],    
\end{aligned}    
\end{equation}    
and the determinant of the induced metric on $\partial\mathcal{M}$ gives
\begin{align}
\sqrt{\sigma}=\left(\frac{\tilde{L}\cos w}{\sin w}\right)^{d-2}\sqrt{1-m_zT_{0}^{d}\sin^{d}w\sin^{2}\theta}\sin^{d-3}\theta\,\sqrt{\widetilde{\chi}}.   
\end{align}

Then to linear order in $m_z$, the variation of the surface term in the entropy functional \eqref{surface term-1} for the excited state from the vacuum is explicitly given as
\begin{align}
\Delta S_{\textrm{surf}}
&=-\frac{\Omega_{d-3}}{4G_{d+1}}\int_{w=w_{\epsilon}}d\theta
\left(\frac{\tilde{L}\cos w}{\sin w}\right)^{d-2}
\frac{m_zT_{0}^{d}\sin^{d}w}{2}\sin^{d-3}\theta \notag\\
&\quad\times\left[
\frac{4\lambda L^{2}\left(d-4+d\cos^{2}\theta\right)}
{\tilde{L}\cos w(d-2)(d-3)}-\frac{4L^{4}\mu\left(d-8+3d\cos^{2}\theta\right)}
{\tilde{L}^{3}\cos^{3}w(d-2)(d-5)}
\left(3\sin^{2}w-1\right)\right].
\end{align}
Note that the boundary divergent terms cancel against the vacuum contribution, while
the parameter $m_z$ affects only terms of order $\mathcal{O}(w_\epsilon^2)$ or higher,
with $w_\epsilon=u_\epsilon/T_0\simeq\epsilon/T_0$, which vanish at the boundary. Thus, at least to the linear order perturbation, the surface term does not contribute to the variation of HTEE for the hyperbolic entangling surface in cubic Lovelock gravity.

Finally, substituting eqs.~\eqref{bar R}, \eqref{bar L4}, \eqref{square h},
\eqref{delta R-1} and \eqref{delta L4} into eq.~\eqref{HEE-1}, the variation
$\Delta S=S-\left.S\right|_{m_z=0}$ up to the order $O(m_z)$ is
\begin{align}
\Delta S
&=\frac{\Omega_{d-3}}{4G_{d+1}}
\int_{\epsilon/T_{0}}^{\pi/2}dw\int_{0}^{\pi}d\theta
\frac{\tilde{L}^{d-1}m_zT_{0}^{d}\cos^{d-2}w\sin w
\sin^{d-3}\theta}{2}\notag\\
&\quad\times\left\{
(\cos^{2}w-\sin^{2}w\cos^{2}\theta-\sin^{2}\theta)
\left(1-2\lambda\frac{L^{2}}{\tilde{L}^{2}}\left(\frac{d-1}{d-3}\right)
-\frac{3L^{4}\mu}{\tilde{L}^{4}}\left(\frac{d-1}{d-5}\right)\right)
\right. \notag\\
&\quad\left.
+\frac{2}{\tilde{L}^{2}}
\bigg((2d+1)\cos^{2}w-(d^{2}+d-1)\cos^{2}w\cos^{2}\theta-d+2\bigg)
\right. \notag\\
&\quad\left.
\times\left(\frac{2L^{2}\lambda}{(d-2)(d-3)}
+\frac{6L^{4}\mu}{\tilde{L}^{2}(d-2)(d-5)}\right)
\right\},\notag\\
&=-\frac{\Omega_{d-3}\tilde{L}^{d-1}m_zT_{0}^{d}}{8G_{d+1}}
\frac{\sqrt{\pi}\,\Gamma\!\left(\frac{d-2}{2}\right)}
{(d^{2}-1)\Gamma\!\left(\frac{d-1}{2}\right)}\notag\\
&\quad\times
\left(1-2\lambda\frac{L^{2}}{\tilde{L}^{2}}\left(\frac{d-1}{d-3}\right)
-\frac{3L^{4}\mu}{\tilde{L}^{4}}\left(\frac{d-1}{d-5}\right)
+\frac{4L^{2}\lambda}{\tilde{L}^{2}(d-3)}
+\frac{12L^{4}\mu}{\tilde{L}^{4}(d-5)}\right),\notag \\
 &=-\frac{\Omega_{d-2}\tilde{L}^{d-1}m_zT_{0}^{d}}{8G_{d+1}(d^2-1)}
 \left(1-2\lambda\frac{L^{2}}{\tilde{L}^{2}}
 -\frac{3L^{4}\mu}{\tilde{L}^{4}}\right),\label{eq:variance S}
\end{align}
in which $\Omega_{d}$ is the volume of the $d$-dimensional unit
sphere, and in the last step we have used the relation $
\Omega_{d-2}/\Omega_{d-3}=\sqrt{\pi}\Gamma\left(\frac{d-2}{2}\right)/\Gamma\left(\frac{d-1}{2}\right)$
to simplify the expression. Moreover, in deriving eq.~\eqref{eq:variance S}, we have used the
fact that the bulk surface extremizing the functional does not change at this
order: the shape variation is protected under the low-energy excitation up to the order $O(m_z)$~\cite{Guo:2013aca}.
As a consistent check, we reproduce the result given in~\cite{Li:2025tud}
with the parameters $\lambda=0,\mu=0$. 

On the modular Hamiltonian side, ref.~\cite{Bhattacharya_2013} proposed a
universal first law like relation between the energy variation and the
entanglement entropy variation for small subsystems. This relation defines an
entanglement temperature proportional to the inverse radius of the spherical
entangling surface. It was shown in~\cite{Guo:2013aca} that the same entanglement
temperature is obtained in higher derivative gravity for the $d=4$ and $d=6$
CFTs dual to Lovelock gravity. The higher curvature interactions that appear in
the entropy variation cancel out of the entanglement temperature. Ref.~\cite{Li:2025tud} extended this first law like relation to the TEE, in which the effective entanglement temperature is inversely proportional to the
temporal size of the interval. We now examine the corresponding relation for the HTEE in the context of Lovelock gravity. The essential step is the calculation of the boundary stress tensor associated with low-energy excitation in Lovelock gravity on asymptotically AdS spacetime.

Considering the coordinate transformation
$\frac{d\zeta}{\zeta}=\frac{\sqrt{1+m_zz^{d}}}{z}dz$ in eq.~\eqref{induced metric-1}, up to first order,
we obtain the perturbation of the metric in the FG gauge as
\begin{align}
       \delta g_{\tau\tau}&=m_z\zeta^{d}\left(\frac{1-d}{d}\right),\notag\\
       \delta g_{(d-1)(d-1)}&=-\frac{m_z\zeta^d}{d} .
   \end{align}
When $m_{\textrm{max}}=3$, the boundary stress tensor can be obtained from \cite{Guo:2013aca}
\begin{align}
    T_{\mu\nu}
    &=-\frac{1}{8\pi G_{d+1}}\left[
    K_{\mu\nu}-K\frac{\tl^2}{\zeta^2}h_{\mu\nu}
    +\frac{2\lambda L^2}{(d-2)(d-3)}
    \left(3J_{\mu\nu}-J\frac{\tl^2}{\zeta^2}h_{\mu\nu}\right)\right. \notag\\
    &\quad\left.
    -\frac{3\mu L^4}{(d-2)(d-3)(d-4)(d-5)}
    \left(5P_{\mu\nu}-P\frac{\tl^2}{\zeta^2}h_{\mu\nu}\right)
    +\frac{d-1}{L''}\frac{\tl^2}{\zeta^2} h_{\mu\nu}
    \right],
\end{align}
in which $h_{\mu\nu}$ is the induced metric on the AdS boundary and
  \begin{align}
       J_{\mu\nu}&=\frac{1}{3}\left(2KK_{\mu\rho}K^\rho_\nu+K_{\rho\sigma}K^{\rho\sigma}K_{\mu\nu}-2K_{\mu\rho}K^{\rho\sigma}K_{\sigma\nu}-K^2K_{\mu\nu}\right),\\
       J&=h^{\mu\nu}J_{\mu\nu}
   \end{align}
and
\begin{align}
P_{\mu\nu}
&=\frac{1}{5}\Big[\Big(K^4-6K^2K_{\rho\sigma}K^{\rho\sigma}
+8KK_{\rho\sigma}K_{\lambda}^{\sigma}K^{\lambda\rho}
-6K_{\rho\sigma}K^{\sigma\lambda}K_{\lambda\kappa}K^{\kappa\rho} \notag\\
&\qquad
+3(K_{\rho\sigma}K^{\rho\sigma})^2\Big)K_{\mu\nu}
-(4K^3-12KK_{\rho\sigma}K^{\rho\sigma}
+8K_{\rho\sigma}K_{\lambda}^{\sigma}K^{\lambda\rho})K_{\mu\lambda}K_{\nu}^{\lambda} \notag\\
&\qquad
-24KK_{\mu\lambda}K^{\lambda\kappa}K_{\kappa\rho}K_{\nu}^{\rho}
+12(K^2-K_{\rho\sigma}K^{\rho\sigma})K_{\mu\lambda}K^{\lambda\kappa}K_{\kappa\nu} \notag\\
&\qquad
+24K_{\mu\lambda}K^{\lambda\kappa}K_{\kappa\rho}K^{\rho\sigma}K_{\sigma\nu}\Big],
\end{align}
where $\mu,\nu,\rho,\sigma,\lambda,\kappa=0,\dots,d-1$ and $K_{\mu\nu}$ is the extrinsic curvature
of the conformal boundary with $K=K_{\mu\nu}h^{\mu\nu}$. The length $L''$ satisfies
\begin{align}
\frac{1}{L''}=\frac{1}{L'}-\frac{3\mu L^4}{5\tl^3},
\qquad
L'=\frac{3\tilde{L}^3}{3\tilde{L}^2-2L^2\lambda}.    
\end{align}
Using the formula of the extrinsic curvature in eq.~\eqref{K-conformal}, the variation of the boundary stress tensor is
\begin{align}
   \Delta T_{(d-1)(d-1)}&=-\frac{\tilde{L}^{d-1}m_z}{16\pi G_{d+1}}\left(1-2\lambda\frac{L^{2}}{\tilde{L}^{2}}-\frac{3L^{4}\mu}{\tilde{L}^{4}}\right)\\
    \Delta T_{\tau\tau}&=(d-1)\Delta T_{(d-1)(d-1)}.\label{energy momentum tensor}
\end{align}
Given eqs.~\eqref{eq:variance S} and \eqref{energy momentum tensor},
the timelike entanglement temperature in higher curvature gravity is also
proportional to the inverse radius of the hyperbolic entangling surface and
reproduces the result of~\cite{Li:2025tud}. This occurs because the
higher curvature factor in eq.~\eqref{eq:variance S} is exactly canceled by the
corresponding factor in eq.~\eqref{energy momentum tensor}. However, as noted in~\cite{Li:2025tud}, the entanglement temperature is not universal.
The quantity we are more concerned with is the modular Hamiltonian,
which allows the timelike entanglement first law to be formulated
precisely. 

By implementing a double Wick rotation, ref. \cite{Li:2025tud}
obtained the exact modular Hamiltonian for the hyperbolic subsystem
in the boundary CFT:
\begin{align} \label{var H}
\Delta \langle H \rangle & =2\pi\int_{\mathcal{D}'}d\tau d^{d-2}\mathbf{x}\,
\frac{\left(T_{0}^{2}-\tau^{2}-\mathbf{x}^{2}\right)}{2T_{0}}\Delta T_{(d-1)(d-1)}.
\end{align}
Note that the form of the modular Hamiltonian remains identical to that in Einstein gravity, as the assumption of a conformally flat boundary ensures it is unaffected by higher curvature interactions in the bulk.
Comparing with eq.~\eqref{eq:variance S} yields
\begin{align}
\Delta\langle H\rangle & =\frac{2\pi T_{0}^{d}\Omega_{d-2}}{\left(d^{2}-1\right)}\Delta T_{(d-1)(d-1)}=\Delta S,
\end{align}
which implies that, for low-energy thermal excitations, the timelike entanglement
first law also holds in $(d+1)$-dimensional cubic Lovelock gravity. In
section~\ref{sec5} we will show that the same matching persists for
Lovelock gravity of arbitrary order within the normalizable perturbative
sector around the AdS spacetime background.

\section{Linearized field equations in Lovelock gravity}\label{sec4}
With the timelike entanglement first law in Lovelock gravity now preliminarily established, it is natural to investigate the corresponding field equations in this framework. Let us consider a general $(d+1)$-dimensional Lovelock gravity, whose
action is given in \eqref{action}.
The field equations of this theory can be written as

\begin{equation}
\sum_{m=0}c_{m}L^{2m-2}\mathcal{E}_{AB}^{(m)}=0,\label{L:e.o.m}
\end{equation}
where we have defined the tensor
\begin{align}
    \mathcal{E}_{AB}^{(m)}&=\frac{1}{\sqrt{-g}}\frac{\delta(\sqrt{-g}\mathcal{L}_{m})}{\delta g^{AB}}\notag\\
    &=-\frac{1}{2^{m+1}}g_{AC}\delta_{BB_{1}\dots B_{2m}}^{CA_{1}\dots A_{2m}}\prod _{q=1}^m R^{B_{2q-1}B_{2q}}_{\phantom{A_{2q-1}A_{2q}}A_{2q-1}A_{2q}}.
\label{lovelock-tensor}
\end{align}
As anticipated, since our chosen ground state corresponds to the pure $\textrm{AdS}_{d+1}$ solution in eq.~\eqref{pure AdS-1-1}, the background spacetime is maximally symmetric,
with metric $\bar{g}_{AB}$. The Riemann tensor of such spacetime
is given by
\begin{equation}
\bar{R}_{ABCD}=-\frac{1}{\tilde{L}^{2}}\left(\bar{g}_{AC}\bar{g}_{BD}-\bar{g}_{AD}\bar{g}_{BC}\right).\label{m.s.s.}
\end{equation}
Evaluated on the background metric, eq.~\eqref{lovelock-tensor} is reformulated  as 
\begin{align}
    \bar{\mathcal{E}}_{AB}^{(m)}&=\frac{(-)^{m+1}}{2\tl^{2m}}\frac{d!}{(d-2m)!}\bar{g}_{AB}.
\end{align}
For Einstein gravity, the maximally symmetric spacetime satisfying
eq.~\eqref{m.s.s.} is a solution of the theory provided $\tilde{L}$ is
related to $L$ through 
\begin{align}
\frac{d(d-1)}{\tilde{L}^{2}}=\frac{c_{0}}{L^{2}},\qquad c_{1}=1.   
\end{align}

Now consider a small perturbation on the background AdS$_{d+1}$ spacetime, i.e., $g_{AB}=\bar{g}_{AB}+\delta g_{AB}$, with $\delta g_{AB}\ll1$
for all $A,B=0,\ldots,d$.  
Using 
\begin{align}
&g_{AC}\delta\bigg(\delta_{BB_{1}\dots B_{2m}}^{CA_{1}\dots A_{2m}}\prod _{q=1}^m R^{B_{2q-1}B_{2q}}_{\phantom{A_{2q-1}A_{2q}}A_{2q-1}A_{2q}}\bigg)\notag\\
    &=2^{m}\frac{(d-2)!}{(d-2m)!}\frac{(-)^{m-1}}{\tl^{2m-2}}\left[g_{AB}\delta R-2\left(\frac{d}{\tl^2}\delta g_{AB}+\delta R_{AB}\right)\right],
\end{align}
then the linearized version of tensor $\mathcal{E}_{AB}^{(m)}$ is given by
\begin{align}
    \delta \mathcal{E}_{AB}^{(m)}=&\frac{(-)^{m+1}}{2\tl^{2m}}\frac{d!}{(d-2m)!}\delta g_{AB}\notag\\
    &+\frac{(d-2)!}{(d-2m)!}\frac{(-)^{m}}{\tl^{2m-2}}\left[\frac{1}{2}g_{AB}\delta R-\left(\frac{d}{\tl^2}\delta g_{AB}+\delta R_{AB}\right)\right].
\end{align}
Using the equations of motion of pure AdS$_{d+1}$ background 
\begin{align}
    \sum_{m\ge 0}^{[\frac{d+1}{2}]}c_m L^{2m-2}\frac{(-)^m d!}{(d-2m)!}\frac{1}{\tl^{2m}}=0,\label{e.o.m of lovelock}
\end{align}
we can obtain the linearized field equations
\begin{align}\label{linear e.o.m} 
    &\sum_{m=0}c_{m}L^{2m-2}\delta \mathcal{E}_{AB}^{(m)}\notag\\
    &=\sum_{m=1}c_m L^{2m-2}m\frac{(d-2)!}{(d-2m)!}\frac{(-)^{m+1}}{\tl^{2m-2}}\left(\delta R_{AB}-\frac{1}{2}g_{AB}\delta R+\frac{d}{\tl^2}\delta g_{AB}\right)\\\notag
    & =\left(1-\sum_{m=2}m\lambda_{m}f_{\infty}^{m-1}\right)\delta G_{AB}=0,   
\end{align}
where in the second line we have used the notation of section \ref{lovelock gravity},
namely $\tilde{L}^{2}=L^{2}/f_{\infty}$, $\lambda_{m}=(-)^{m}\frac{(d-2)!}{(d-2m)!}c_{m}$, and $\delta G_{AB} =\delta R_{AB}-\frac{1}{2}g_{AB}\delta R+\frac{d}{\tl^2}\delta g_{AB}$ for the linearized Einstein equation.
Equation \eqref{linear e.o.m} shows that on a maximally symmetric background
(such as AdS spacetime), the correction arising from Lovelock
terms only appears in the overall factor of the equation of motion,
while all kinetic terms coincide with the ones from Einstein gravity.
Equivalently, the contribution of Lovelock terms to the linearized
field equations amounts only to a renormalization of the effective
Newtonian constant
\begin{equation}
G_{\textrm{eff}}^{-1}=G_{d+1}^{-1}\left(1-\sum_{m=2}m\lambda_{m}f_{\infty}^{m-1}\right).\label{effective G}
\end{equation}
This implies that, around a maximally symmetric background, Lovelock terms do
not change the tensor structure of the linearized field equations; they only rescale
the overall gravitational coupling constant. The property is special to Lovelock gravity
and is the reason why a first law matching can reduce to a comparison of the
same coupling dependent factor.

Now let us revisit the results of section~\ref{sec3}. For
low-energy thermal excitations in cubic Lovelock gravity, eq.~\eqref{eq:variance S} gives
\begin{align}
\Delta S & =-\frac{\Omega_{d-2}\tilde{L}^{d-1}m_zT_{0}^{d}}{8G_{d+1}(d^2-1)}
 \left(1-2\lambda\frac{L^{2}}{\tilde{L}^{2}}
 -\frac{3L^{4}\mu}{\tilde{L}^{4}}\right),\label{eq:variance S-2}\notag\\
 & =\Delta S_{\textrm{Einstein}}\left(1-2\lambda\frac{L^{2}}{\tilde{L}^{2}}-\frac{3L^{4}\mu}{\tilde{L}^{4}}\right),
\end{align}
where $\Delta S_{\textrm{Einstein}}$ is the variation
of HTEE in Einstein gravity~\cite{Li:2025tud}.
Using the stress tensor of the boundary CFT dual to cubic Lovelock
gravity, the variation of the modular Hamiltonian for the hyperbolic subsystem
after double Wick rotation is
\begin{align}
\Delta \langle H \rangle & =2\pi\int_{\mathcal{D}'}d\tau d^{d-2}\mathbf{x}\,
\frac{\left(T_{0}^{2}-\tau^{2}-\mathbf{x}^{2}\right)}{2T_{0}}\Delta T_{(d-1)(d-1)},\notag\\
 & =\Delta \langle H \rangle_{\textrm{Einstein}}\left(1-2\lambda\frac{L^{2}}{\tilde{L}^{2}}-\frac{3L^{4}\mu}{\tilde{L}^{4}}\right),
\end{align}
where $\Delta \langle H \rangle_{\textrm{Einstein}}$ is the variation of modular Hamiltonian of the boundary CFT with an Einstein gravity dual~\cite{Li:2025tud}.
The overall coefficient obtained from the boundary stress tensor is exactly
the same as the one in eq.~\eqref{eq:variance S-2}. Since the timelike
entanglement first law has already been proven in Einstein gravity~\cite{Li:2025tud}, the common coefficient implies that the first law also holds for low-energy thermal excitations in $(d+1)$-dimensional cubic Lovelock gravity. Moreover, for $m_{\textrm{max}}=3$ this higher derivative coefficient is the same as that renormalizes the effective Newtonian constant in
eq.~\eqref{effective G}. Therefore, apart from the modified effective gravitational coupling constant, the linearized perturbations behave as in Einstein gravity. In a maximally symmetric background, the equivalence between the timelike
entanglement first law and the linearized field equations in Einstein gravity
therefore extends to cubic Lovelock gravity for the low-energy thermal excitations
considered here. In the next section, we will establish the corresponding statement
for Lovelock gravity of arbitrary order within the normalizable perturbative
sector around the AdS background.

\section{General analysis}\label{sec5}
In this section, we extend the cubic analysis to Lovelock gravity of arbitrary
order by keeping the same controlled setting: perturbations around the AdS
vacuum, a fixed conformally flat boundary geometry, FG gauge and a hyperbolic
timelike subregion with a local modular Hamiltonian. Thus, the result proved
below is not a statement about arbitrary holographic states. It is a statement
about the normalizable perturbative sector in which the double Wick rotated
description and the hyperbolic modular Hamiltonian are both under analytic
control. We continue to work with the hyperbolic subsystem $\m$ embedded in a
double Wick-rotated spacetime, the unperturbed ground state metric takes the
form
\begin{align}
    ds^2&=\frac{\tl^2}{z^2}(dz^2+d\tau^2+d\textbf{x}^2-dx_{d-1}^2), \quad \textbf{x}\in \mathbb{R}^{d-2}.
\end{align}
Consider a general metric perturbation subject to the transverse traceless (TT) gauge constraints ($\bar{g}^{AB}\delta g_{AB}=\bar{\nabla}^{A}\delta g_{AB}=0$) as well as the FG gauge conditions ($\delta g_{z\mu}=\delta g_{zz}=0$). The near-boundary asymptotic expansion of the boundary metric is given by
\begin{align}
    \delta \tilde{h}_{\mu\nu}&=\delta \tilde{h}_{\mu\nu}^{(0)}+\dots+z^d\delta\tilde{h}_{\mu\nu}^{(d)}+z^d\log{z^2}\delta \hat{h}_{\mu\nu}^{(d)}+\dots,
\end{align}
where the logarithmic coefficient $\delta \hat{h}_{\mu\nu}^{(d)}$ captures the conformal anomaly. For a conformally flat boundary, the subleading terms $\delta h^{(n)}_{\mu\nu}$ ($n<d$) are entirely functionally dependent on the non-normalizable mode $\delta\tilde{h}^{(0)}_{\mu\nu}$~\cite{deHaro:2000vlm,Skenderis:2000in}. 

Crucially, the relative entropy quantifies the divergence between two distinct density matrices defined on the identical CFT Hilbert space, which dictates that the background manifold must be held rigid. Imposing this physical requirement sets $\delta \tilde{h}_{\mu\nu}^{(0)}=0$. By extension, all subleading terms $n<d$ and the anomaly-driven logarithmic term strictly vanish. Therefore, the physical normalizable perturbation reduces to the compact form
\begin{align}
    \delta \tilde{h}_{\mu\nu}&=z^d\delta \tilde{h}^{(d)}_{\mu\nu}+\dots .\label{FG perturbation}
\end{align}

\subsection{Stress tensor in general Lovelock gravity}
To explicitly evaluate $\Delta \langle H \rangle$, we must derive the holographic stress tensor in general Lovelock gravity, as prescribed by eq.~\eqref{delta H}. A prerequisite for this computation, and for establishing a well posed variational principle, is the explicit construction of the boundary surface term and its associated counter-terms, in which the generalized surface term can be formulated by introducing an an auxiliary field~\cite{Deruelle:2009zk}. In this formalism, the generalized Gibbons-Hawking-York (GHY) boundary term takes the integral form
\begin{align}
    \tilde{I}_{\text{surf}} &= \frac{1}{4\pi G_{d+1}}\int d^dx\sqrt{-h} \Psi ^{AB}K_{AB},
\end{align}
where $h_{\mu\nu}$ denotes the induced metric on the spacetime boundary. The conjugate auxiliary field $\Psi^{AB}$ is defined as the variation of the Lagrangian with respect to the bulk Riemann tensor, projected onto the normal directions:
\begin{align}
    \Psi^{AB} &= \frac{\partial \rl}{\partial R_{ACBD}}n_Cn_D.
\end{align}
As shown in \eqref{K-conformal}, the extrinsic curvature does not contain a $z$-component; thus, $K_{AB}=\delta^{\mu}{}_A\delta^{\nu}{}_B K_{\mu\nu}$.
The underlying dynamics are governed by the Lovelock Lagrangian density $\rl$, which is given by 
\begin{align}
    \mathcal{L} &= \sum_{m\ge0}^{[\frac{d+1}{2}]}c_m L^{2m-2}\mathcal{L}_{m}(R)\notag\\
    &= \sum_{m\ge0}^{[\frac{d+1}{2}]}c_m L^{2m-2}\frac{1}{2^m}\delta_{A_1 A_2 \cdots A_{2m-1} A_{2m}}^{B_1 B_2 \cdots B_{2m-1} B_{2m}}\prod_{q=1}^m g^{A_{2q-1}C_{2q-1}}g^{A_{2q}D_{2q}}R_{C_{2q-1}D_{2q}B_{2q-1}B_{2q}}.
\end{align}

Our next objective is to translate the bulk curvature degrees of freedom into boundary geometric quantities. To systematically decompose the intrinsic bulk curvature into the intrinsic and extrinsic curvatures of the boundary, we invoke the Gauss-Codazzi equation~\cite{Misner:1973prb}:
\begin{align}
    \hat{R}_{\mu\nu\rho\sigma}(h) &= h_{\mu}^{A}h_{\nu}^{B}h_{\rho}^{C}h_{\sigma}^{D}R_{ABCD}[g]+K_{\mu\rho}K_{\nu\sigma}-K_{\mu\sigma}K_{\nu\rho}.
\end{align}
In our holographic setup, we evaluate this expression near a conformally flat boundary. Consequently, the intrinsic boundary Riemann tensor $\hat{R}_{\mu\nu\rho\sigma}(h)$ identically vanishes. Furthermore, aligning the cutoff boundary with a constant slice $z=\epsilon$ (parameterized by coordinates $X=(z,x^\mu)$) yields a trivial boundary metric and a unit normal vector pointing purely in the $z$-direction. This orthogonal foliation simplifies the projection of the bulk Riemann tensor to
\begin{align}
    R_{\mu\nu\rho\sigma}^{\parallel }[g] &= K_{\mu\sigma}K_{\nu\rho}-K_{\mu\rho}K_{\nu\sigma}.
\end{align}

With the curvature decomposed, we must now tackle the highly non-trivial tensor contractions embedded within the Lovelock action. By applying the following combinatorial identity for the generalized Kronecker delta function, which effectively isolates the normal components,
\begin{align}
    &\delta_{A_1 A_2 \cdots A_{2m-1} A_{2m}}^{B_1 B_2 \cdots B_{2m-1} B_{2m}}g^{A_{1}C_{1}}g^{A_{2}D_{2}}\frac{1}{2}\left(\delta^{[\rho}_{C_{1}}\delta^{z]}_{D_{2}}\delta ^{[\nu}_{B_{1}}\delta^{z]}_{B_{2}}+\delta^{[\nu}_{C_{1}}\delta^{z]}_{D_{2}}\delta ^{[\rho}_{B_{1}}\delta^{z]}_{B_{2}}\right)n_{z}n_{z}K_{\rho\nu}\notag\\
    &= K^\rho_\nu\delta_{\rho\rho_1 \cdots  \rho_{2m-2}}^{\nu\nu_1 \cdots  \nu_{2m-2}},
\end{align}
we can explicitly perform the contraction between the auxiliary field and the extrinsic curvature. Unpacking the series step by step, we find
\begin{align}
    &\frac{\partial \mathcal{L}}{\partial R_{ABCD}}n_{B}n_{D}\delta^\mu{}_{A}\delta^\nu{}_{C}K_{\mu\nu}\notag\\
    &= \sum_{m=1}^{[\frac{d+1}{2}]}c_m L^{2m-2}\frac{1}{2^m}K^\rho_\nu\delta_{\rho\rho_1 \cdots \rho_{2m-2}}^{\nu \nu_1\cdots \nu_{2m-2} }m\prod_{q=1}^{m-1}h^{\rho_{2q-1}\lambda_{2q-1}}h^{\rho_{2q}\kappa_{2q}}R^{\parallel }_{\lambda_{2q-1}\kappa_{2q}\nu_{2q-1}\nu_{2q}} \notag\\
    &= \sum_{m=1}^{[\frac{d+1}{2}]}c_mL^{2m-2}\frac{(-2)^{m-1}m}{2^m}\delta_{\nu_0 \cdots \nu_{2m-2}}^{\mu_0 \cdots \mu_{2m-2}}K^{\nu_0}_{\mu_0}\dots K^{\nu_{2m-2}}_{\mu_{2m-2}}.
\end{align}
Finally, integrating this contracted quantity over the boundary manifold yields the closed-form expression for the generalized GHY surface term:
\begin{align}
    \tilde{I}_{\text{surf}} &= \frac{1}{4\pi G_{d+1}}\int d^dx\sqrt{-h}\sum_{m=1}^{[\frac{d+1}{2}]}c_mL^{2m-2}\frac{(-1)^{m-1}m}{2}\delta_{\nu_0 \cdots \nu_{2p-2}}^{\mu_0 \cdots \mu_{2p-2}}K^{\nu_0}_{\mu_0}\dots K^{\nu_{2m-2}}_{\mu_{2m-2}}.
\end{align}

However, the generalized GHY term constructed above does not immediately serve as the correct Myers surface term. The crux of this disparity lies in the fact that, upon employing the Gauss-Codazzi decomposition in Lovelock gravity, the auxiliary field implicitly acquires a dependence on the extrinsic curvature, $\Psi^{AB}(K_{\rho\sigma})$. Consequently, taking the full variation of the generalized GHY term generates anomalous contributions:
\begin{align}
    \delta\tilde{I}_{\text{surf}}\propto \int_{\partial M} \left( \Psi ^{AB}\delta K_{AB}+K_{AB}\delta \Psi ^{AB}+\dots \right).
\end{align}
On the contrary, integrating the bulk variation by parts produces a specific boundary flux:
\begin{align}
    \delta I_{\text{bulk}}\propto \int _{\partial M}\Psi ^{AB}\delta K_{AB}+\dots .
\end{align}
To guarantee a well-posed Dirichlet variational principle, the true surface term must be engineered such that its variation exactly absorbs the bulk flux:
\begin{align}
    \delta I_{\text{surf}}\propto \int_{\partial M}\Psi ^{AB}\delta K_{AB}+\dots .
\end{align}
This canonical requirement establishes a fundamental differential relation between the physical Myers surface term and the generalized GHY term:
\begin{align}
    K_{AB}\frac{\partial I_{\text{surf}}}{\partial K_{AB}} &= \tilde{I}_{\text{surf}}.
\end{align}
For the $m$-th order Lovelock Lagrangian evaluated on a conformally flat boundary, this scaling relation simplifies to~\cite{Liu:2017kml}
\begin{align}
    I_{\text{surf}}^{(m)} &= \frac{\tilde{I}_{\text{surf}}^{(m)}}{2m-1}.
\end{align}
Combining the exact Myers boundary term with the bulk action, we evaluate the full bare action as
\begin{align}
    I &= I_{\text{bulk}}+I_{\text{surf}}\notag\\
    &= \frac{1}{16\pi G_{d+1}}\int d^{d+1}x \frac{\tl^{d+1}}{z^{d+1}}\sqrt{-\tilde{g}}\sum_{m\ge 0}^{[\frac{d+1}{2}]}c_m L^{2m-2}\frac{(d+1)!}{(d+1-2m)!}\frac{(-1)^{m}}{\tl^{2m}}\notag\\
    &\quad +\frac{1}{8\pi G_{d+1}}\int d^dx \frac{\tl^d}{z^d}\sqrt{-\tilde{h}}\sum_{m=1}^{[\frac{d+1}{2}]}c_mL^{2m-2}
    \frac{(-1)^{m-1}m}{\tl^{2m-1}}\frac{d!}{(2m-1)(d-2m+1)!}.
\end{align}

In holographic renormalization, the appropriate counterterm is obtained by isolating the UV divergences in the limit $z\to\epsilon$. This yields
\begin{align}
    I_{\text{ct}} &= -\frac{1}{16\pi G_{d+1}}\int d^dx\frac{\tl^{d+1}}{\epsilon^d}\sqrt{-\tilde{h}}\notag\\
    &\quad \times \left[c_0\frac{1}{dL^2}+\sum_{m=1}^{[\frac{d+1}{2}]}\tl^{-2m}c_m L^{2m-2}\left(\frac{(d+1)!}{(d+1-2m)!}\frac{(-1)^{m}}{d}+\frac{2(-1)^{m-1}md!}{(2m-1)(d-2m+1)!}\right)\right].
\end{align}
By imposing the background equations of motion, i.e., eq.~\eqref{e.o.m of lovelock}, the counterterm simplifies to
\begin{align}
    I_{\text{ct}} &= \frac{1}{16\pi G_{d+1}}\int d^dx\sqrt{-h}\sum_{m=1}^{[\frac{d+1}{2}]}\tl^{1-2m}c_mL^{2m-2}
    \frac{(-1)^{m}2m(d-1)!}{(2m-1)(d-2m)!}.
\end{align}
With the total renormalized action formally defined, the boundary stress tensor can be extracted via functional differentiation with respect to the induced metric:
\begin{align}
    \tilde{T}^{\mu\nu} &= \frac{2}{\sqrt{-h}}\frac{\delta I_{\text{tot}}}{\delta h_{\mu\nu}}\notag\\
    &= \frac{2}{\sqrt{-h}}\frac{\delta( I+I_{\text{ct}})}{\delta h_{\mu\nu}}\notag\\
    &= 2\pi^{\mu\nu}+T^{\mu\nu}_{\text{ct}}.
\end{align}
The contribution arising purely from the counterterm evaluates to
\begin{align}
    \tilde{T}^{\mu\nu}_{\text{ct}} &= \frac{1}{16\pi G_{d+1}}h^{\mu\nu}\sum_{
    m=1}^{[\frac{d+1}{2}]}\tl^{1-2m}c_mL^{2m-2}\frac{(-1)^{m}2m(d-1)!}{(2m-1)(d-2m)!}.
\end{align}
Although explicit evaluation of the bare action $I$ has been achieved, computing the canonical momentum $\pi^{\mu\nu}$ directly from it is practically intractable. To systematically circumvent this, we adopt the technique developed in \cite{Deruelle:2009zk}.

In the ADM decomposition, the bulk action can be reformulated in a Hamiltonian form as
\begin{align}
    I_{\text{bulk}} &= \frac{1}{16\pi G_{d+1}}\int d^{d}xdz N\sqrt{-h}\tilde{\mathcal{L}}(h_{\mu\nu},K_{\mu\nu},R^{\parallel}_{\mu\nu\rho\sigma}[g],R^{\parallel}_{\mu\nu\rho\mathbf{n}}[g]),
\end{align}
where $R^{\parallel}_{\mu\nu\rho\mathbf{n}}[g]=\nabla_\mu K_{\nu\rho}-\nabla_\nu K_{\mu\rho}$ encapsulates the higher-derivative nature of the metric along the normal direction, and $N=\frac{\tl}{z}$. The corresponding canonical momentum conjugate to the induced metric $h_{\mu\nu}$ is then formally defined as
\begin{align}
    \pi^{\mu\nu} &= \frac{\delta I_{\text{bulk}}}{\delta \partial_{z}h_{\mu\nu}} \notag\\
    &= \frac{\sqrt{-h}}{32\pi G_{d+1}}\frac{\partial\mathcal{L}}{\partial K_{\mu\nu}}.
\end{align}
Performing a Legendre transformation, we construct the boundary Hamiltonian density. For Lovelock gravity, a remarkable simplification occurs, yielding~\cite{Teitelboim:1987zz}
\begin{align}
    \mathcal{H} &= K_{\mu\nu}\pi^{\mu\nu}-\mathcal{L} \notag\\
    &= -\mathcal{L}(R^{\parallel}_{\mu\nu\rho\sigma}[g]),
\end{align}
where the effective reduced Lagrangian $\mathcal{L}(R^{\parallel}_{\mu\nu\rho\sigma}[g])$ is given by
\begin{align}
    \mathcal{L}(R^{\parallel}_{\mu\nu\rho\sigma}[g]) &= \sum _{m\ge0}^{[\frac{d+1}{2}]}c_mL^{2m-2}(-1)^m\delta_{\nu_0 \cdots \nu_{2m-1}}^{\mu_0 \cdots \mu_{2m-1}}K^{\nu_0}_{\mu_0}\dots K^{\nu_{2m-1}}_{\mu_{2m-1}},
\end{align}
with $R^{\parallel}_{\mu\nu\rho\sigma}[g]$ being the intrinsic curvature projected onto the boundary. 

Analogous to the surface term, the exact relationship between the reduced Hamiltonian and the true canonical momentum is established as
\begin{align}
    \pi^{\mu\nu(m)} & = \frac{1}{2(2m-1)}\frac{\partial \mathcal{H}^{(m)}}{\partial K_{\mu\nu}},
\end{align}
Evaluating this explicitly on the background geometry, the mixed tensor components of the momentum are
\begin{align}
    \pi^\mu_\nu &= -\frac{1}{16\pi G_{d+1}}\sum _{m=1}^{[\frac{d+1}{2}]}c_mL^{2m-2}\frac{m(-1)^m}{(2m-1)}\delta_{\nu \nu_1\cdots \nu_{2m-1}}^{\mu\mu_1 \cdots \mu_{2m-1}}K^{\nu_1}_{\mu_1}\dots K^{\nu_{2m-1}}_{\mu_{2m-1}} \notag\\
    &= -\frac{1}{16\pi G_{d+1}}\sum _{m=1}^{[\frac{d+1}{2}]}c_mL^{2m-2}\frac{m(-1)^m}{(2m-1)}\frac{1}{\tl^{2m-1}}
    \frac{(d-1)!}{(d-2m)!}\delta^\mu_\nu.
\end{align}
To compute the response of the stress tensor to metric perturbations, we take the variation of this canonical momentum, which isolates the linear order extrinsic dynamics:
\begin{align}
    \delta \pi ^\mu_\nu &= -\frac{1}{16\pi G_{d+1}}\sum _{m=1}^{[\frac{d+1}{2}]}c_mL^{2m-2}\frac{m(-1)^m}{\tl^{2m-2}}\frac{(d-2)!}{(d-2m)!}(\delta K\delta ^\mu_\nu-\delta K^\mu_\nu),
\end{align}    
and the variation of the full momentum tensor is $\delta \pi^{\mu\nu}=\pi^{\mu}_\rho\delta h^{\rho\nu} + h^{\rho\nu}\delta\pi^\mu_\rho$. Within the standard holographic renormalization scheme, the contribution from the first term precisely cancels the divergent boundary counterterms, leaving only the physically finite part sourced by $\delta\pi^\mu_\rho$.

Consequently, taking into account that the metric perturbation strictly satisfies the FG gauge as formulated in eq.~\eqref{FG perturbation}, the variation of the holographic stress tensor reduces to
\begin{align}
    \Delta \tilde{T}_{\mu\nu} &= 2h_{\nu\rho}\delta \pi _\mu^\rho \notag\\
    &= -\frac{d}{16\pi G_{d+1}}\frac{\tl}{z^{d-2}}\delta \tilde{h}_{\mu\nu}^{(d)}\sum _{m=1}^{[\frac{d+1}{2}]}c_mL^{2m-2}\frac{m(-1)^m}{\tl^{2m-2}}\frac{(d-2)!}{(d-2m)!}.
\end{align}
To extract the universal physical quantities defined on the CFT boundary, we apply the standard conformal scaling limit $z \to 0$:
\begin{align}
    \Delta T_{\mu\nu} &= \lim _{z\to 0}\left(\frac{\tl}{z}\right)^{d-2}\Delta \tilde{T}_{\mu\nu} \notag\\
    &= -\frac{d\tl^{d-1}}{16\pi G_{d+1}}\sum _{m=1}^{[\frac{d+1}{2}]}m(-1)^m c_m\left(\frac{L}{\tl}\right)^{2m-2}\frac{(d-2)!}{(d-2m)!}\delta \tilde{h}_{\mu\nu}^{(d)}.
\end{align}
Finally, the variation of the modular Hamiltonian in eq.~\eqref{delta H} takes the explicit closed form as
\begin{align}
    \Delta \langle H \rangle  &= \frac{\pi}{T_0}\int_{\mathcal{D}'} d\tau d^{d-2}\mathbf{x}\,
    (T_0^2-\tau^2-\mathbf{x}^2)\Delta T_{(d-1)(d-1)} \notag\\
    &= -\tl^{d-1}\sum _{m=1}^{[\frac{d+1}{2}]}m(-1)^m c_m\left(\frac{L}{\tl}\right)^{2m-2}\frac{(d-2)!}{(d-2m)!}\notag\\
    &\quad \times\frac{d}{16 G_{d+1}T_0}\int_{\mathcal{D}'} d\tau d^{d-2}\mathbf{x}\,
    (T_0^2-\tau^2-\mathbf{x}^2)\delta \tilde{h}_{(d-1)(d-1)}^{(d)}\notag\\
    &=-\sum _{m=1}^{[\frac{d+1}{2}]}m(-1)^m c_m\left(\frac{L}{\tl}\right)^{2m-2}\frac{(d-2)!}{(d-2m)!}\Delta \langle H \rangle_{\textrm{Einstein}}\notag\\
    &=\left(1-\sum _{m=2}^{[\frac{d+1}{2}]}m\lambda_m f_\infty^{m-1}\right)\Delta \langle H \rangle_{\textrm{Einstein}}.\label{H in arbitrary}
\end{align}
where $\Delta \langle H \rangle_{\textrm{Einstein}}$ is the variation of the modular Hamiltonian for the boundary CFT with an Einstein's gravity dual~\cite{Li:2025tud}.
Eq.~\eqref{H in arbitrary} is the central matching relation on the
modular Hamiltonian side. The coefficient multiplying the Einstein gravity result is the same combination that appears in the effective Newtonian constant
in eq.~\eqref{effective G}. In the following subsections, we will show that the JM variation gives precisely the same factor, ,while the additional contribution from the generalized surface term vanishes in the conformal limit.

\subsection{The extremal configuration of the surface functional}
This section is devoted to the derivation of the variation of TEE within the framework of general Lovelock gravity. Although previous studies have shown that the extremal surface retains its spherical geometry for $m=2$ and $m=3$~\cite{Hung:2011xb,Guo:2013aca}, this result does not immediately generalize to an arbitrary order. Consequently, we first show that, within the class of rotationally symmetric surfaces relevant for the ball-shaped boundary subregion in double Wick rotation space, the spherical embedding remains an extremum for arbitrary Lovelock order.

The HTEE in Lovelock gravity can be expressed as
\begin{align}
    S &= \frac{1}{4G_{d+1}}\int_{\m} d^{d-1}x \sqrt{\gamma}\mathcal{S} + S_{\text{surf}},
\end{align}
where $\mathcal{S} = 1 + \sum_{m=2}^{[\frac{d+1}{2}]} m c_m L^{2m-2} \mathcal{L}_{m-1}(\r)$. The associated surface term can be systematically constructed via the introduction of an auxiliary field, and we define the generalized surface term as
\begin{align}
    \tilde{S}_{\text{surf}} &= \frac{1}{G_{d+1}}\int_{\partial \m} d^{d-2}x \sqrt{\sigma}\psi^{ij}\mathcal{K}_{ij},
\end{align}
where $\psi^{ij} = \frac{\partial \s}{\partial \r_{ikjl}}n_kn_l$. Exploiting the spherical symmetry of the boundary $\partial \m$, we can generally parameterize the geometry as follows:
\begin{align}
    \xi(u) = f(u/T_0)\cos(u/T_0), \quad z(u) = f(u/T_0)\sin(u/T_0), \quad \epsilon \leq u\leq \frac{\pi}{2}T_0,\label{parameterization}
\end{align}
where $\xi^2 = \tau^2 + \mathbf{x}^2$, with $\mathbf{x}\in \mathbb{R}^{d-2}$ and $x_{d-1}=\text{const}$. In this framework, the boundary $\partial \m$ is situated at the constant-$u$ hypersurface $u=u_\epsilon$, dictating that the unit normal vector points purely along the $u$-direction.

Evaluating the tensor contractions yields
\begin{align}
    &\frac{\partial \s}{\partial \r_{iuju}}n_{u}n_{u}\delta^a{}_i\delta^b{}_j\k_{ab} \notag\\
    &= \sum_{m=2}^{[\frac{d+1}{2}]}mc_mL^{2m-2}\frac{1}{2^{m-1}}\k^b_a\delta_{b b_1\cdots b_{2m-4}}^{a a_1\cdots a_{2m-4}}(m-1)\prod_{k=1}^{m-2}\r_{\phantom{\nu_{2k-1}\nu_{2k}}a_{2k-1}a_{2k}}^{\parallel b_{2k-1}b_{2k}}[\gamma],
\end{align}
where $\r_{\phantom{\nu_{2k-1}\nu_{2k}}a_{2k-1}a_{2k}}^{\parallel b_{2k-1}b_{2k}}[\gamma]$ denotes the curvature tensor of the extremal surface projected onto the boundary $\partial \m$. By invoking the Gauss-Codazzi equations, these components can be decomposed in terms of the intrinsic and extrinsic curvatures on the entangling surface. The generalized surface term thus reads~\cite{Liu:2017kml}
\begin{align}
    \tilde{S}_{\text{surf}} &= \frac{1}{G_{d+1}}\int_{\partial\m}d^{d-2}x\sqrt{\sigma} \sum_{m=2}^{[\frac{d+1}{2}]}mc_mL^{2m-2}\frac{1}{2^{m-1}}\k^b_a\delta_{bb_1 \cdots b_{2m-4}}^{aa_1 \cdots a_{2m-4}}(m-1) \notag\\
    &\quad \times \prod_{k=1}^{m-2}\left(\r_{\phantom{\nu_{2k-1}\nu_{2k}}a_{2k-1}a_{2k}}^{\partial b_{2k-1}b_{2k}}[\sigma]-2\k^{b_{2k-1}}_{a_{2k-1}}\k^{b_{2k}}_{a_{2k}}\right),
\end{align}
where $\r_{\phantom{\nu_{2k-1}\nu_{2k}}a_{2k-1}a_{2k}}^{\partial b_{2k-1}b_{2k}}[\sigma]$ represents the intrinsic Riemann curvature of $\partial \m$. 

The actual surface term is systematically reconstructed by integrating the differential relation $\k_{ab}\frac{\partial S_{\text{surf}}}{\partial \k_{ab}} = \tilde{S}_{\text{surf}}$, which gives
\begin{align}
    S_{\text{surf}}(\r^{\partial}_{abcd},\k_{ab}) &= \int_0^1 \tilde{S}_{\text{surf}}(\r^{\partial}_{abcd},s\k_{ab})\frac{ds}{s} \notag\\
    &= \frac{1}{G_{d+1}}\int_0^1 ds
    \int_{\partial\m}d^{d-2}x\sqrt{\sigma}
    \sum_{m=2}^{[\frac{d+1}{2}]}mc_mL^{2m-2}
    \frac{(m-1)}{2^{m-1}}\delta_{b b_1\cdots b_{2m-4}}^{a a_1\cdots a_{2m-4}} \notag\\
    &\quad \times
    \prod_{k=1}^{m-2}\Big(
    \r_{\phantom{\nu_{2k-1}\nu_{2k}}a_{2k-1}a_{2k}}^{\partial b_{2k-1}b_{2k}}[\sigma]
    -2s^2\k^{b_{2k-1}}_{a_{2k-1}}\k^{b_{2k}}_{a_{2k}}
    \Big)\k^b_a.
\end{align}
Analytical integration gives the explicit form:
\begin{align}
    S_{\text{surf}}
    &= \frac{1}{G_{d+1}}\int_{\partial\m}d^{d-2}x\sqrt{\sigma}
    \sum_{m=2}^{[\frac{d+1}{2}]}mc_mL^{2m-2}
    \frac{(m-1)}{2^{m-1}}\delta_{b b_1\cdots b_{2m-4}}^{aa_1 \cdots a_{2m-4}} \notag\\
    &\quad \times \k^b_a
    \sum_{l=0}^{m-2}\frac{(m-2)!(-2)^l}{(m-2-l)!l!(2l+1)}
    \prod_{k=1}^{m-2-l}
    \r^{\partial b_{2k-1}b_{2k}}_{\phantom{a_{2k-1}b_{2k}}a_{2k-1}a_{2k}}[\sigma] \notag\\
    &\quad \times
    \prod_{i=1}^l
    \k_{a_{2(m-2-l)+2i-1}}^{b_{2(m-2-l)+2i-1}}
    \k_{a_{2(m-2-l)+2i}}^{b_{2(m-2-l)+2i}}. \label{surface of HTEE}
\end{align}
The induced metric on the extremal surface naturally admits a warped product structure, taking the form
\begin{align}
    d\hat{s}^2 &= \frac{\tl^2}{z^2}\bigg(\left[(\partial_u f)^2 + f^2/T_0^2\right]du^2 + f^2\cos(u/T_0)^2 d\Omega_{d-2}^2\bigg) \notag\\
    &= ds_X^2 + e^{2F}ds_Y^2,
\end{align}
where the warp factor is determined by $F = \log\Big(\frac{\tl \xi}{z}\Big)$. In this expression, we have explicitly decomposed the metric into a one-dimensional base sector $ds_X^2 = \frac{\tl^2}{z^2}[(\partial_u f)^2 + f^2/T_0^2] du^2$ and a $(d-2)$-dimensional angular sector $ds_Y^2 = \tilde{\sigma}_{ab}d\theta^a d\theta^b$, with indices $a,b=1,\dots,d-2$. 

By exploiting this warped geometry, the non-vanishing components of the intrinsic Riemann curvature tensor can be systematically evaluated. The mixed components involving the $u$-direction are given by~\cite{Hung:2011xb}
\begin{align}
    \r_{uaub}[\gamma] &= -e^{2F}(\nabla_u\nabla_u F + \partial_u F\partial_u F)\tilde{\sigma}_{ab} \notag\\
    &= -e^{2F}\gamma_{uu}\frac{1}{\sqrt{\gamma_{uu}}}\partial_u\left(\sqrt{\gamma_{uu}}\gamma^{uu}\dot{F}\right)
    \tilde{\sigma}_{ab} - e^{2F}\gamma_{uu}\dot{F}^2\tilde{\sigma}_{ab},
\end{align}
where the overdot denotes the derivative with respect to $u$, i.e., $\dot{F} = \partial_u F$. Furthermore, the purely angular components yield
\begin{align}
    \r^{\parallel}_{abcd}[\gamma] &= e^{2F}\r^{Y}_{abcd} + \gamma^{uu}\dot{F}^2e^{4F}\big(\tilde{\sigma}_{bc}\tilde{\sigma}_{ad}-\tilde{\sigma}_{bd}\tilde{\sigma}_{ac}\big) \notag\\
    &= \r^{\partial}_{abcd}[\sigma] + \gamma^{uu}\dot{F}^2e^{4F}\big(\tilde{\sigma}_{bc}\tilde{\sigma}_{ad}-\tilde{\sigma}_{bd}\tilde{\sigma}_{ac}\big).
\end{align}
Given these explicit curvature components, the systematic expansion of the Lovelock Lagrangian density can now be carried out. By separating the index contractions along the $u$-direction from the purely angular ones, the Euler density decomposes as
\begin{align}
    \mathcal{L}_{m-1}(\r)
    &= \frac{1}{2^{m-1}}
    \delta_{i_1 i_2 \cdots i_{2m-3} i_{2m-2}}
    ^{j_1 j_2 \cdots j_{2m-3} j_{2m-2}}
    \r^{i_1 i_2}_{\phantom{j_1 j_2}j_1 j_2}\cdots
    \r^{i_{2m-3} i_{2m-2}}
    _{\phantom{j_{2p-3} j_{2p-2}}j_{2m-3} j_{2m-2}}
    \notag\\
    &= \frac{4(m-1)}{2^{m-1}}
    \delta_{a_1 a_2 \cdots a_{2m-4} a_{2m-3}}
    ^{b_1 b_2 \cdots b_{2m-4} b_{2m-3}}
    \r^{u a_1}_{\phantom{j_1 j_2}u b_1}
    \prod_{k=1}^{m-2}
    \r^{\parallel a_{2k} a_{2k+1}}
    _{\phantom{j_{2k-1} j_{3p-2}}b_{2k} b_{2k+1}}
    + \mathcal{L}_{m-1}(\r^{\parallel}_{abcd}).
\end{align}
It is important to note that the purely angular term $\mathcal{L}_{m-1}(\r^{\parallel}_{abcd})$ vanishes identically for dimensional reasons whenever $2m>d$.

Isolating the dynamics of the embedding, we focus strictly on the terms containing the derivative of $\gamma_{uu}$, as these are responsible for generating the dangerous second derivatives of $f$. This specific contribution is given by
\begin{align}
    \mathcal{L}_{m-1}^{\text{sec}}
    &= \frac{4(m-1)}{2^{m-1}}
    \delta_{a_1 a_2 \cdots a_{2m-4} a_{2m-3}}
    ^{b_1 b_2 \cdots b_{2m-4} b_{2m-3}}
    \left(-\frac{1}{\sqrt{\gamma_{uu}}}
    \partial_u\left(\sqrt{\gamma_{uu}}\gamma^{uu}\right)\dot{F}\delta^{a_1}_{b_1}\right)
    \notag\\
    &\quad\times
    \prod_{k=1}^{m-2}
    \r^{\parallel a_{2k} a_{2k+1}}
    _{\phantom{\nu_{2k-1} \nu_{3p-2}}b_{2k} b_{2k+1}}
    \notag\\
    &= \frac{4(m-1)(d-2m+2)}{2^{m-1}}
    \delta_{a_2 \cdots a_{2m-4} a_{2m-3}}
    ^{b_2 \cdots b_{2m-4} b_{2m-3}}
    \left(-\frac{1}{\sqrt{\gamma_{uu}}}
    \partial_u\left(\sqrt{\gamma_{uu}}\gamma^{uu}\right)\dot{F}\right) \notag\\
    &\quad\times
    \prod_{k=1}^{m-2}
    \r^{\parallel a_{2k} a_{2k+1}}
    _{\phantom{\nu_{2k-1} \nu_{3p-2}}b_{2k} b_{2k+1}}.
\end{align}
The corresponding contribution to the TEE from the above bulk term is
\begin{align}
    S^{\text{sec}}
    &= \frac{1}{4G_{d+1}}\sum_{m=2}^{[\frac{d+1}{2}]}
    mc_mL^{2m-2}\int_{\m}d^{d-1}x
    \sqrt{\sigma}\sqrt{\gamma_{uu}} \notag\\
    &\quad \times
    \frac{4(m-1)(d-2m+2)}{2^{m-1}}
    \delta_{a_1 \cdots a_{2m-5} a_{2m-4}}
    ^{b_1 \cdots b_{2m-5} b_{2m-4}}
    \left(-\frac{1}{\sqrt{\gamma_{uu}}}
    \partial_u\left(\sqrt{\gamma_{uu}}\gamma^{uu}\right)\dot{F}\right) \notag\\
    &\quad \times
    \prod_{k=1}^{m-2}
    \r^{\parallel a_{2k-1} a_{2k}}
    _{\phantom{\nu_{2k-1} \nu_{3p-2}}b_{2k-1} b_{2k}}.
\end{align}
We can perform a combinatorial simplification on the tensor contractions by substituting the expanded angular curvature:
\begin{align}
    &\delta_{a_1 \cdots a_{2m-5} a_{2m-4}}^{b_1 \cdots b_{2m-5} b_{2m-4}}\prod_{k=1}^{m-2} \r^{\parallel a_{2k-1} a_{2k}}_{\phantom{\nu_{2k-1} \nu_{3p-2}}b_{2k-1} b_{2k}} \notag\\
    &= \delta_{a_1 \cdots a_{2m-5} a_{2m-4}}
    ^{b_1 \cdots b_{2m-5} b_{2m-4}}
    \prod_{k=1}^{m-2}\left[
    \r^{\partial a_{2k-1}a_{2k}}
    _{\phantom{a_{2k}a_{2k+1}}b_{2k-1}b_{2k}}[\sigma]
    -2\gamma^{uu}\dot{F}^2
    \delta^{a_{2k-1}}_{b_{2k-1}}\delta^{a_{2k}}_{b_{2k}}
    \right] \notag\\
    &= \sum_{l=0}^{m-2}
    \frac{(m-2)!(-2)^l(\dot{F})^{2l}(\gamma^{uu})^l}{(m-2-l)!l!}
    \prod_{k=1}^{m-2-l}
    \r^{\partial b_{2k-1}b_{2k}}
    _{\phantom{a_{2k-1}b_{2k}}a_{2k-1}a_{2k}}[\sigma] \notag\\
    &\quad \times
    \prod_{i=1}^l
    \delta_{a_{2(m-2-l)+2i-1}}^{b_{2(m-2-l)+2i-1}}
    \delta_{a_{2(m-2-l)+2i}}^{b_{2(m-2-l)+2i}}
    \delta_{a_1 \cdots a_{2m-5} a_{2m-4}}
    ^{b_1 \cdots b_{2m-5} b_{2m-4}},
\end{align}
and further applying the dimensional reduction identity for generalized Kronecker deltas functions:
\begin{align}
    &\prod_{i=1}^l\delta_{a_{2(m-2-l)+2i-1}}^{b_{2(m-2-l)+2i-1}}\delta_{a_{2(m-2-l)+2i}}^{b_{2(m-2-l)+2i}} \delta_{a_1 \cdots a_{2m-5} a_{2m-4}}^{b_1 \cdots b_{2m-5} b_{2m-4}} \notag\\
    &= \delta^{b_1\dots b_{2m-4-2l}}_{a_1\dots a_{2m-4-2l}}\frac{(d-2-2(m-2-l))!}{(d-2-(2m-4))!} \notag\\
    &= \delta^{b_1\dots b_{2m-4-2l}}_{a_1\dots a_{2m-4-2l}}\prod_{i=1}^{2l}\left(d-2-(2m-4)+i\right).
\end{align}
By isolating an individual $(m,l)$-th mode from the summations, the bulk integrand simplifies to
\begin{align}
    S^{\text{sec}}_{(m)(l)}
    &= \frac{4(m-1)(d-2m+2)}{2^{m-1}}
    \frac{(m-2)!(-2)^l(\dot{F})^{2l}\dot{F}(\gamma^{uu})^l}
    {(m-2-l)!l!}
    \prod_{i=1}^{2l}\Big[\left(d-2-(2m-4)+i\right)\Big] \notag\\
    &\quad \times
    \sqrt{\gamma_{uu}}\left(-\frac{1}{\sqrt{\gamma_{uu}}}
    \partial_u\frac{1}{\sqrt{\gamma_{uu}}}\right)
    \prod_{k=1}^{m-2-l}
    \r^{\partial b_{2k-1}b_{2k}}
    _{\phantom{a_{2k-1}b_{2k}}a_{2k-1}a_{2k}}[\sigma]
    \delta^{b_1\dots b_{2m-4-2l}}_{a_1\dots a_{2m-4-2l}} \notag\\
    &= C_{m,l}\left(-(\gamma^{uu})^l\partial_u\frac{1}{\sqrt{\gamma_{uu}}}\right),
\end{align}
where all purely geometric and numerical factors have been absorbed into the coefficient $C_{m,l}$:
\begin{align}
    C_{m,l}
    &= \frac{4(m-1)(d-2m+2)}{2^{m-1}}
    \frac{(m-2)!(-2)^l(\dot{F})^{2l}\dot{F}}{(m-2-l)!l!}
    \prod_{i=1}^{2l}\Big[(d-2-(2m-4)+i)\Big] \notag\\
    &\quad \times
    \prod_{k=1}^{m-2-l}
    \r^{\partial b_{2k-1}b_{2k}}
    _{\phantom{a_{2k-1}b_{2k}}a_{2k-1}a_{2k}}[\sigma]
    \delta^{b_1\dots b_{2m-4-2l}}_{a_1\dots a_{2m-4-2l}}.
\end{align}

Turning to the surface term, we must verify whether it precisely cancels the term with derivative of $\gamma_{uu}$. Given the unit normal vector $n_i = -\sqrt{\gamma_{uu}}\delta_{ui},i=u,1,\dots,d-2$, the components of the extrinsic curvature are evaluated as
\begin{align}\label{exkab}
    \k_{ab} &= -\frac{1}{N}e^{2F}\dot{F}\tilde{\sigma}_{ab}, \notag\\
    \k_b^a &= -\frac{1}{N}\dot{F}\delta_b^a,
\end{align}
where $N = \sqrt{\gamma_{uu}}$. Substituting eq.~\eqref{exkab} into the expression for the generalized boundary term, i.e., eq.~\eqref{surface of HTEE}, the corresponding $(m,l)$-th contribution from the surface reads
\begin{align}
    S_{\text{surf}(m)(l)}
    &= \frac{4(m-1)}{2^{m-1}}
    \delta_{b b_1 \cdots b_{2m-4}}^{a a_1 \cdots a_{2m-4}} \notag\\
    &\quad \times \k^b_a
    \frac{(m-2)!(-2)^l}{(m-2-l)!l!(2l+1)}
    \prod_{k=1}^{m-2-l}
    \r^{\partial b_{2k-1}b_{2k}}
    _{\phantom{a_{2k-1}b_{2k}}a_{2k-1}a_{2k}}[\sigma] \notag\\
    &\quad \times
    \prod_{i=1}^l
    \k_{a_{2(m-2-l)+2i-1}}^{b_{2(m-2-l)+2i-1}}
    \k_{a_{2(m-2-l)+2i}}^{b_{2(m-2-l)+2i}} \notag\\
    &= \frac{-4(m-1)}{2^{m-1}}
    \frac{(m-2)!(-2)^l\dot{F}^{2l}\dot{F}}
    {(m-2-l)!l!(2l+1)}
    \frac{(\gamma^{uu})^l}{\sqrt{\gamma_{uu}}}
    \prod_{k=1}^{m-2-l}
    \r^{\partial b_{2k-1}b_{2k}}
    _{\phantom{a_{2k-1}b_{2k}}a_{2k-1}a_{2k}}[\sigma] \notag\\
    &\quad \times
    \delta^{a_1\dots a_{2m-4-2l}}_{b_1\dots b_{2m-4-2l}}
    (d-2m+2)\prod_{i=1}^{2l}(d-2-(2m-4)+i) \notag\\
    &= -C_{m,l}\frac{1}{(2l+1)}\frac{(\gamma^{uu})^l}{\sqrt{\gamma_{uu}}}.
\end{align}
The exact cancellation becomes transparent when we utilize the algebraic derivative identity
\begin{align}
    \frac{1}{2l+1}\partial_u\left(\frac{(\gamma^{uu})^l}{\sqrt{\gamma_{uu}}}\right) &= (\gamma^{uu})^l\partial_u\frac{1}{\sqrt{\gamma_{uu}}},
\end{align}
combined with the boundary evaluation via the fundamental theorem of calculus:
\begin{align}
    \int_\epsilon^{\frac{\pi}{2}T_0} du \partial_u W &= -W_\epsilon.
\end{align}
Remarkably, the boundary term exactly absorbs the first derivative contributions of $\gamma_{uu}$ originating from the bulk action. Consequently, the total variation of the HTEE is stripped of any second order derivatives of $f$, leaving a functional that depends on $f$ only through its logarithmic derivative:
\begin{align}
    S &= S\left(\frac{d\log{f}}{du}, u\right).
\end{align}
Applying the variational principle to this well-posed action yields a first-integral condition for $f$:
\begin{align}
    \frac{d}{du}\left(\frac{\partial \s}{\partial \big(\frac{d\log{f}}{du}\big)}\right) &= 0,
\end{align}
which is solved by $f=\text{const}$ with the boundary conditions imposed above. This establishes, within the present ansatz, that the extremal surface remains spherical in general Lovelock gravity, thereby preserving its maximally symmetric geometry.

Using the boundary condition $f(u/T_0)=T_0$, the parameterization in eq.~\eqref{parameterization} becomes
\begin{align}
    \xi(u) = T_0\cos(u/T_0), \quad z(u) = T_0\sin(u/T_0), \quad \epsilon \leq u\leq \frac{\pi}{2}T_0.
\end{align}
Then the induced metric on the extremal surface becomes
\begin{align}
    d\hat{s}^2 &= \gamma_{ij}dx^idx^j \notag\\
    &= \frac{\tl^2}{(T_0^2-\xi^2)}\left(\frac{T_0^2}{T_0^2-\xi^2}d\xi^2+\xi^2d\Omega_{d-2}^2\right).
\end{align}
Furthermore, as shown in eq.~\eqref{Kfor extremal}, the vanishing of the extrinsic curvature implies that the Wald entropy coincides with the JM entropy in this case.

\subsection{Timelike entanglement entropy in general Lovelock gravity}
To explicitly display the perturbation on the induced metric, we reformulate it in the following form
\begin{align}
    d\hat{s}^2&=\gamma_{ij}dx^idx^j\notag\\
    &=\frac{\tl^2}{z^2}(\tilde{\gamma}_{ij}+\tilde{g}_{zz}\frac{x_ix_j}{z^2})dx^idx^j,\quad i,j=0,\dots,d-2,
\end{align}
where $\tilde{\gamma}_{ij}$ is the pullback of the boundary metric $\tilde{h}_{\mu\nu}$ to the extremal surface and $\tilde{g}_{zz}=1$ for the AdS background. The inverse metric is $\gamma^{ij}=\frac{z^2}{\tl^2}\frac{\left(T_0^2\tilde{\gamma}^{ij}-x^ix^j\right)}{T_0^2}$. The perturbation on the background gives $\delta \gamma_{ij}=\delta \gamma^{\mathrm{pb}}_{ij}$, where $\delta \gamma^{\mathrm{pb}}_{ij}$ denotes the pullback of $\delta h_{\mu\nu}$.

Evaluating the Euler density on this perturbed metric yields the corresponding
background value and its first order variation,
\begin{align}
 \bar{\mathcal{L}}_{m-1} &= \frac{(d-1)!}{(d+1-2m)!}\left(-\frac{1}{\tl^2}\right)^{m-1},\\
    \delta\mathcal{L}_{m-1} &= \left(-\frac{1}{\tl^2}\right)^{m-2}\frac{(m-1)(d-3)!}{(d+1-2m)!}\delta \r.
\end{align}
And the variation of the HTEE can be systematically expanded as
\begin{align}
    \Delta S &= \frac{1}{4G_{d+1}}\int_{\m}d^{d-1}x\frac{1}{2}\sqrt{\gamma}\gamma^{ij}\delta \gamma_{ij}+\Delta S_{m-1}+\Delta S_{\text{surf}},
\end{align}
in which
\begin{align}
    \Delta S_{m-1}&=\frac{1}{4G_{d+1}}\int_{\m}d^{d-1}x\delta(\sqrt{\gamma}\sum_{m=2}^{[\frac{d+1}{2}]}mc_mL^{2m-2} \rl_{m-1})\notag\\
    &=\frac{1}{4G_{d+1}}\int_{\m}d^{d-1}x\sqrt{\gamma}
    \sum_{m=2}^{[\frac{d+1}{2}]}mc_mL^{2m-2}\bigg[\left(\frac{1}{2}\gamma^{ij}\mathcal{L}_{m-1}-\frac{(-)^{m-2}}{\tl^{2m-4}}\frac{(m-1)(d-3)!}{(d+1-2m)!}\r^{ij}\right)
    \delta \gamma_{ij}\notag\\&+
    \frac{(-)^{m-2}}{\tl^{2m-4}}\frac{(m-1)(d-3)!}{(d+1-2m)!}\gamma^{ij}\delta \r_{ij}\bigg].
\end{align}
For the perturbations specified above, neither the anomalous bulk curvature
variation nor the explicit surface term contributes to the physical variation
of the HTEE. This conclusion relies on two facts that  are specific to the present
setup. First, by the Palatini identity, the bulk variation of the Ricci tensor
organizes into a total derivative and reduces to a boundary flux. Second, for
normalizable FG perturbations with fixed conformally flat boundary metric, the
near-boundary scaling on the hyperbolic cutoff slice
$T_0^2-\xi^2\to\epsilon^2$ suppresses this flux and the generalized surface
term to $\mathcal{O}(\epsilon^2)$. We do not claim that the same suppression
holds for non-normalizable sources, non-conformally-flat boundary metrics or
non-hyperbolic entangling regions.

To demonstrate this explicitly, we isolate the bulk total derivative term and rewrite it as a boundary integral:
\begin{align}
    \Delta S^{\rm D}_{m-1}&=\frac{1}{4G_{d+1}}\sum_{m=2}^{[\frac{d+1}{2}]}mc_mL^{2m-2} \frac{(-1)^{m-2}}{\tl^{2m-4}}\frac{(m-1)(d-3)!}{(d+1-2m)!}\int_{\m}d^{d-1}x\sqrt{\gamma} \gamma^{ij}\delta \r_{ij}\notag\\
    &=\frac{1}{4G_{d+1}}\sum_{m=2}^{[\frac{d+1}{2}]}mc_mL^{2m-2} \frac{(-1)^{m-2}}{\tl^{2m-4}}\frac{(m-1)(d-3)!}{(d+1-2m)!}\int_{\partial\m}d^{d-2}x\sqrt{\sigma}\notag\\
    &\quad \times \left(\sigma^{ab}\delta \Gamma^\xi_{ab}-\gamma^{\xi\xi}\delta \Gamma^a_{a\xi}\right)n_\xi.
\end{align}
To facilitate the evaluation of the radial limits, we employ the coordinate parameterization defined in eq.\eqref{Kee section} in the Appendix.

Evaluating the geometric flux explicitly yields
\begin{align}
    &\left(\sigma^{ab}\delta \Gamma^\xi_{ab}-\gamma^{\xi\xi}\delta \Gamma^a_{a\xi}\right)n_\xi\notag\\
    &=\frac{(T_0^2-\xi^2)}{\tl T_0}\left(-\sigma^{ab}\partial_\xi\delta\sigma_{ab}+\frac{T_0^2}{\xi(T_0^2-\xi^2)}\sigma^{ab}\delta \sigma_{ab}+\nabla^a\delta \gamma_{a\xi}+\frac{(d-2)(T_0^2-\xi^2)}{\xi\tl^2 }\delta \gamma_{\xi\xi}\right).
\end{align}
Here, the integral of the transverse divergence $\nabla^a\delta \gamma_{a\xi}$ evaluates to zero identically, as the spatial slice of the entangling surface $\partial \m$ is a closed sphere without boundary. 

To extract the physical contribution, we take the UV limit $T_0^2-\xi^2 = \epsilon^2 \to 0$. From the asymptotic expansion eq.~\eqref{FG perturbation}, the divergence of background metric dictates that the boundary volume form scales as $\sqrt{\sigma} \sim \mathcal{O}(\epsilon^{-(d-2)})$. In contrast, the physical metric perturbations carry a scaling weight of $\mathcal{O}(\epsilon^{d})$. Unpacking these constraints for the specific tensor components gives:
\begin{align}
    \sqrt{\sigma}=\mathcal{O}(\epsilon^{-(d-2)}),\quad \sigma^{ab}\delta\sigma_{ab}&=\mathcal{O}(\epsilon^d),\quad \delta \gamma_{\xi\xi}=\mathcal{O}(\epsilon^{d-2}).
\end{align}
Multiplying the volume measure by the order of the integrand, the overall behavior of the bulk variation is strictly bounded by
\begin{align}
    \Delta S_{m-1}^{\rm D}=\mathcal{O}(\epsilon^2).
\end{align}
Consequently, this term entirely decouples from the physical HTEE when the cutoff is sent to zero.

Next, we consider the variation of the surface term eq.~\eqref{surface of HTEE}:
\begin{align}
    \Delta S_{\text{surf}} &= \frac{1}{G_{d+1}}\int_{\partial\m}d^{d-2}x\sum_{m=2}^{[\frac{d+1}{2}]}mc_mL^{2m-2}\frac{1}{2^{m-1}}(m-1) \sum_{l=0}^{m-2}\frac{(m-2)!(-2)^l}{(m-2-l)!l!(2l+1)}\notag\\
    &\quad \times \delta \left[\delta_{b \cdots b_{2m-4}}^{a \cdots a_{2m-4}}\sqrt{\sigma}\prod_{k=1}^{m-2-l}\r^{\partial b_{2k-1}b_{2k}}_{\phantom{a_{2k-1}b_{2k}}a_{2k-1}a_{2k}}[\sigma]\prod_{i=1}^l\k_{a_{2(m-2-l)+2i-1}}^{b_{2(m-2-l)+2i-1}}
    \k_{a_{2(m-2-l)+2i}}^{b_{2(m-2-l)+2i}}\k^b_a\right].
\end{align}
Since $\r^{\partial b_{2k-1}b_{2k}}_{\phantom{a_{2k-1}b_{2k}}a_{2k-1}a_{2k}}[\sigma]$ denotes the intrinsic curvature of the sphere modified by the conformal factor, it can be expressed as
\begin{align}
    \r^{\partial b_{2k-1}b_{2k}}_{\phantom{a_{2k-1}b_{2k}}a_{2k-1}a_{2k}}[\sigma]&=\frac{(T_0^2-\xi^2)}{\tl^2\xi^2}\r^{Y b_{2k-1}b_{2k}}_{\phantom{a_{2k-1}b_{2k}}a_{2k-1}a_{2k}}\notag\\
    &=\frac{(T_0^2-\xi^2)}{\tl^2\xi^2}\frac{1}{T_0^2}\left(\delta ^{b_{2k-1}}_{a_{2k-1}}\delta
    ^{b_{2k}}_{a_{2k}}-\delta ^{b_{2k}}_{a_{2k-1}}\delta
    ^{b_{2k-1}}_{a_{2k}}\right).
\end{align}
Consequently, the variation of this quantity is proportional to the variation of the spherical intrinsic curvature:
\begin{align}
    \delta   \r^{\partial b_{2k-1}b_{2k}}_{\phantom{a_{2k-1}b_{2k}}a_{2k-1}a_{2k}}[\sigma]&=\frac{(T_0^2-\xi^2)}{\tl^2\xi^2}\delta \r^{Y b_{2k-1}b_{2k}}_{\phantom{a_{2k-1}b_{2k}}a_{2k-1}a_{2k}}.
\end{align}
Applying the Leibniz rule, the variation of the integrand can be decomposed into contributions from the metric determinant, the intrinsic curvature, and the extrinsic curvature. The coefficient corresponding to the variation of the metric determinant is given by
\begin{align}
   & \delta_{b \cdots b_{2m-4}}^{a \cdots a_{2m-4}}\prod_{k=1}^{m-2-l}\r^{\partial b_{2k-1}b_{2k}}_{\phantom{a_{2k-1}b_{2k}}a_{2k-1}a_{2k}}[\sigma]\prod_{i=1}^l\k_{a_{2(m-2-l)+2i-1}}^{b_{2(m-2-l)+2i-1}}\k_{a_{2(m-2-l)+2i}}^{b_{2(m-2-l)+2i}}\k^b_a\notag\\
   &=-\left(\frac{T_0}{\tl\xi}\right)^{2l+1}\left(\frac{2 (T_0^2-\xi^2)}{\tl^2\xi^2}\right)^{m-2-l}\frac{1}{T_0^{2(m-l-2)}}\frac{(d-2)!}{(d-2m+1)!},
\end{align}
where we have utilized the explicit expression eq.~\eqref{K-entangling} for the extrinsic curvature. 

The variation associated with the intrinsic curvature evaluates to
\begin{align}
   & \delta_{b \cdots b_{2m-4}}^{a \cdots a_{2m-4}}\delta\left(\prod_{k=1}^{m-2-l}\r^{\partial b_{2k-1}b_{2k}}_{\phantom{a_{2k-1}b_{2k}}a_{2k-1}a_{2k}}[\sigma]\right)\prod_{i=1}^l\k_{a_{2(m-2-l)+2i-1}}^{b_{2(m-2-l)+2i-1}}
   \k_{a_{2(m-2-l)+2i}}^{b_{2(m-2-l)+2i}}\k^b_a\notag\\
   &=-(m-2-l)\delta \r^Y\left(\frac{T_0}{\tl\xi}\right)^{2l+1}\left(\frac{2 (T_0^2-\xi^2)}{\tl^2\xi^2}\right)^{m-2-l}\frac{1}{T_0^{2(m-2)-2l-2}}\frac{(d-4)!}{(d-2m+1)!}.
\end{align}
Similarly, the variation generated by the extrinsic curvature is
\begin{align}
     & \delta_{b \cdots b_{2m-4}}^{a \cdots a_{2m-4}}\prod_{k=1}^{m-2-l}\r^{\partial b_{2k-1}b_{2k}}_{\phantom{a_{2k-1}b_{2k}}a_{2k-1}a_{2k}}[\sigma]\delta\left(\prod_{i=1}^l\k_{a_{2(m-2-l)+2i-1}}^{b_{2(m-2-l)+2i-1}}
     \k_{a_{2(m-2-l)+2i}}^{b_{2(m-2-l)+2i}}\k^b_a\right)\notag\\
     &=(2l+1)\delta\k \frac{(d-3)!}{(d-2m+1)!}\left(\frac{T_0}{\tl\xi}\right)^{2l}\left(\frac{2 (T_0^2-\xi^2)}{\tl^2\xi^2}\right)^{m-2-l}\frac{1}{T_0^{2(m-l-2)}}.
\end{align}
Combining these individual contributions, we obtain
\begin{align}
&\delta \left[\delta_{b \cdots b_{2m-4}}^{a \cdots a_{2m-4}}\sqrt{\sigma}\prod_{k=1}^{m-2-l}\r^{\partial b_{2k-1}b_{2k}}_{\phantom{a_{2k-1}b_{2k}}a_{2k-1}a_{2k}}[\sigma]\prod_{i=1}^l\k_{a_{2(m-2-l)+2i-1}}^{b_{2(m-2-l)+2i-1}}
\k_{a_{2(m-2-l)+2i}}^{b_{2(m-2-l)+2i}}\k^b_a\right]\notag\\
    &=\sqrt{\sigma}\left(\frac{T_0}{\tl\xi}\right)^{2l+1}\left(\frac{2 (T_0^2-\xi^2)}{\tl^2\xi^2}\right)^{m-2-l}\frac{1}{T_0^{2(m-l-2)}}\frac{(d-4)!}{(d-2m+1)!}\notag\\
    &\quad \times \left(-(d-3)(d-2)\frac{1}{2}\sigma^{ab}\delta\sigma_{ab}-(m-2-l)T_0^2\delta \r^Y+(2l+1)(d-3)\frac{\tl\xi}{T_0}\delta\k\right).
\end{align}
And for the intrinsic curvature, the variation yields
\begin{align}
    \delta \r^Y&=\frac{4(d-3)}{T_0^2}\delta \tilde{\sigma}\sim \mathcal{O}(\epsilon^d).
\end{align}
Here $\delta\tilde{\sigma}\equiv\tilde{\sigma}^{ab}\delta\tilde{\sigma}_{ab}$.
Regarding the perturbation of the extrinsic curvature, we expand it as
\begin{align}
    \delta\k&=\delta \sigma^{ab}\k_{ab}+\delta\k_{ab}\sigma^{ab}\notag\\
    &=\frac{T_0}{\tl\xi}\delta \sigma+\delta \k_{ab}\sigma^{ab},
\end{align}
where $\delta\sigma\equiv\sigma^{ab}\delta\sigma_{ab}$. Substituting the following relation
\begin{align}
    \delta \k_{ab}\sigma^{ab}&=\frac{(T_0^2-\xi^2)}{2T_0^2\xi^2}\left(\sigma^{ab}\partial_\xi \delta \sigma_{ab}-2\nabla^a\delta \gamma_{a\xi}\right)-\frac{\delta \gamma_{\xi\xi}(T_0^2-\xi^2)^2}{2T_0^4\xi^4}\frac{(d-2)}{\tl}\frac{T_0}{\xi},
\end{align}
and noting that the total derivative term vanishes upon integration as previously established, it follows that
\begin{align}
    \delta \k\sim \mathcal{O}(\epsilon^d).
\end{align}
In conclusion, since the trace of the metric perturbation, the variation of
the intrinsic curvature, and the perturbation of the extrinsic curvature all
scale as $\mathcal{O}(\epsilon^d)$ for the normalizable perturbations under
consideration, combining them with the divergent conformal factor and the
metric determinant leads to an overall scaling of $\mathcal{O}(\epsilon^2)$ for
the surface term variation. Consequently, within this hyperbolic FG setup, it follows that $\Delta S_{\text{surf}}=\mathcal{O}(\epsilon^2)$, thus the boundary
contribution vanishes in the conformal limit $\epsilon \to 0$.

Finally, the variation of the HTEE is
\begin{align}
    \Delta S&=-\frac{1}{4G_{d+1}}\sum_{m=1}^{[\frac{d+1}{2}]}(-)^m mc_m\left(\frac{L}{\tl}\right)^{2m-2}\frac{(d-2)!}{(d-2m)!}\tl^{d-1}\int_{\m}d^{d-1}x\frac{1}{z^{d-1}}\frac{1}{2}\sqrt{\tilde{\gamma}}
    \gamma^{ij}\delta \gamma_{ij}\notag\\
    &=-\frac{1}{4G_{d+1}}\sum_{m=1}^{[\frac{d+1}{2}]}(-)^m mc_m\left(\frac{L}{\tl}\right)^{2m-2}\frac{(d-2)!}{(d-2m)!}\tl^{d-1}\int_{\m}d^{d-1}x\frac{1}{z^{d}}\frac{(T_0^2\tilde{\gamma}^{ij}-x^ix^j)}{2T_0}\delta \tilde{\gamma}_{ij}\notag\\
    &=-\sum_{m=1}^{[\frac{d+1}{2}]}(-)^m mc_m\left(\frac{L}{\tl}\right)^{2m-2}\frac{(d-2)!}{(d-2m)!}\Delta S_{\textrm{Einstein}}\notag\\
    &=\left(1-\sum_{m=2}^{[\frac{d+1}{2}]}m\lambda_m f_\infty^{m-1}\right)\Delta S_{\textrm{Einstein}}
\end{align}
where $\tilde{\gamma}_{ij}=\text{diag}\{1,\dots,1\}$ is the unperturbed pulled-back metric and $i,j=0,\dots,d-2$.
Observe that this overall coefficient is identical to that of the modular
Hamiltonian eq.~\eqref{H in arbitrary}. Together with the result of Einstein gravity
for hyperbolic regions, this establishes the equivalence between the timelike entanglement first law and
the linearized field equations of Lovelock gravity in the restricted perturbative sector
specified at the beginning of this section.

\section{Conclusions and discussions}\label{sec6}
In the present paper, we have studied the timelike entanglement first law in holographic CFTs whose
bulk dual is governed by Lovelock gravity. The analysis combines the double
Wick rotation construction of timelike entanglement with the JM
entropy functional, which is the correct functional for HEE in Lovelock gravity. For hyperbolic timelike subregions we computed the
linear variation of HTEE and compared it with the variation of the modular
Hamiltonian obtained from the holographic stress tensor. In the explicit
cubic Lovelock example, the higher curvature dependence appears through the
identical overall factor in both quantities, which establishes
$\Delta S=\Delta\langle H\rangle$ for low-energy thermal excitations and reduces to
the Einstein-gravity result when the Lovelock couplings are switched off.

We then extended the analysis to arbitrary order Lovelock gravity around the
AdS vacuum. For normalizable perturbations in FG gauge, the
bulk contribution from the JM functional is proportional to Einstein gravity's result with a Lovelock dependent prefactor. In the hyperbolic cutoff geometry
and for a fixed conformal flat boundary metric, the generalized boundary
term does not contribute in the conformal limit, and the same prefactor
appears in the modular Hamiltonian variation. This is precisely the
coefficient that renormalizes the linearized Lovelock field equations about the maximally symmetric AdS spacetime background. Therefore, within the perturbative and hyperbolic regime specified in this work, the timelike entanglement first law is equivalent to the field equations of Lovelock gravity linearized about the AdS spacetime.

Several qualifications are important. Our argument assumes the double Wick
rotation prescription for HTEE and the analytic continuation of the JM
functional described in section \ref{sec2}. It does not constitute a general
Lorentzian replica derivation of the higher derivative timelike entropy
functional. Moreover, the vanishing of the boundary contribution relies on
normalizability, FG gauge, the hyperbolic slicing and a fixed conformally flat
boundary. These restrictions are part of the statement of the result.

There are several extensions would be worth pursuing. First, it would be important to
derive a fully covariant complex surface prescription for HTEE in general
higher derivative theories. Lovelock gravity is special because the JM
functional depends only on intrinsic curvatures of the hypersurface. More general
higher curvature actions require extrinsic curvature corrections, and the
analytic continuation of these terms to timelike or complex saddles is not
yet understood in a systematic way. A clean derivation, ideally tied to a
Lorentzian replica or Noether charge construction, would clarify which entropy
functional should replace eq.~\eqref{JM} beyond the Lovelock class.

Second, the present work focus on perturbations around the vacuum and uses
hyperbolic regions, where the modular Hamiltonian is local. It would be
interesting to extend the analysis to finite excitations, time-dependent
backgrounds and black hole geometries. In such cases the real and imaginary
parts of HTEE may carry different geometric information, and the role of the
surface term could become more delicate. Understanding whether the
$\mathcal{O}(\epsilon^2)$ suppression found here survives outside the
normalizable hyperbolic sector may provide a sharper diagnostic of complex
extremal surfaces in Lorentzian holography.

Third, the relation between timelike entanglement, pseudo entropy and bulk gravitational dynamics deserves further study. The present result suggests that timelike
modular data can still encode gravitational equations when the bulk dynamics
is modified by higher curvature interactions. It would be natural to ask
whether this statement persists for non-hyperbolic regions, for theories with
matter sources, and for settings in which pseudo entropy is computed by genuinely transition matrix rather than analytically continued density matrix
data. These directions may help identify which parts of the entanglement
gravity correspondence are special to spacelike entanglement and which extend
to the broader class of timelike and complex entanglement observables.

\acknowledgments
We would like to thank Horacio Casini and Tadashi Takayanagi for useful discussions. S.H. was supported by the National Natural
Science Foundation of China (NSFC) under Grant
Nos.~12475053, 12588101 and 12235016, and by the sub-project funding for
``Gravitational Redshift Measurement Scientific Experiment and Frontier
Research in Gravitational Physics'' of the Strategic Priority Research Program
on Space Science of the Chinese Academy of Sciences (XDA30040000,
XDA30030000). J.R.S. was supported by the National Natural Science Foundation
of China (No.~12475069) and the Guangdong Basic and Applied Basic Research
Foundation (No.~2025A1515011321).

\appendix
\section{Extrinsic curvature}
Let $M$ be a $(d+1)$-dimensional smooth manifold with boundary $\partial M$. The coordinates on $M$ are denoted by $X^{A}$ ($A=1,\dots,d+1$), and the intrinsic coordinates restricted to the boundary are $x^\mu$ ($\mu=1,\dots,d$). The embedding of the boundary into the bulk manifold is given by $X^{A}=X^{A}(x^\mu)$, and the corresponding tangent basis vectors to the boundary are $e^{A}_\mu=\frac{\partial X^{A}}{\partial x^\mu}$.

Equipping the manifold $M$ with a bulk metric $g_{AB}$, the induced metric on $\partial M$ is defined as $h_{\mu\nu}=g_{AB}e^{A}_\mu e^{B}_\nu$. The unit normal vector to $\partial M$ with respect to $M$ is denoted by $n^{A}$, which satisfies the normalization condition $g_{AB}n^{A}n^{B}=1$ and the orthogonality condition $g_{AB}n^{A}e^{B}_{\mu}=0$. 

The extrinsic curvature characterizes the bending of the boundary within the bulk, quantified by the directional derivative of the unit normal vector along the tangent basis. Therefore, the extrinsic curvature is obtained by calculating
\begin{equation}
    K_{\mu\nu} = e^{A}_{\mu}e^{B}_{\nu}\nabla_{A}n_{B}.
\end{equation}

If the metric is expressed in the ADM formalism, i.e.,
\begin{align}
    ds^2 &= g_{AB}dX^{A}dX^{B}\notag\\
    &= N^2 d\zeta^2 + h_{\mu\nu}(dx^\mu+N^\mu d\zeta)(dx^\nu+N^\nu d\zeta), \label{ADM}
\end{align}
the tangent basis vectors simplify to $e^{A}_{\mu} = \delta^{A}_{\mu}$, since the coordinates are adapted as $X^{A}=(\zeta,x^\mu)$. In this coordinate system, the covariant components of the unit normal vector are strictly orthogonal to the constant-$\zeta$ slices, taking the form $n_{A}=(\pm N, \vec{0})$, where the sign is chosen depending on whether the normal vector is outward- or inward-pointing. 

Noting that $n_\mu = 0$, the definition of the extrinsic curvature simplifies to $K_{\mu\nu} = -\Gamma^{\zeta}_{\mu\nu}n_{\zeta}$. Expressed in terms of the ADM variables, this yields the rigorous form
\begin{align}
    K_{\mu\nu} &= \pm\frac{1}{2N}(\partial_{\zeta}h_{\mu\nu} - \nabla_\mu N_\nu - \nabla_\nu N_\mu), \label{K}
\end{align}
where $\nabla_\mu$ denotes the covariant derivative strictly compatible with the induced boundary metric $h_{\mu\nu}$.
\subsection{Extrinsic curvature of the conformal boundary}
Considering Lovelock gravity and performing a double Wick rotation, the corresponding metric becomes 
\begin{align}
    ds^2 &= \frac{\tl^2}{z^2}dz^2 + \frac{\tl^2}{z^2}\tilde{h}_{\mu\nu}dx^\mu dx^\nu, \quad \mu=0,\dots,d-1,
\end{align}
where $\tilde{h}_{\mu\nu}=\text{diag}\{1,\dots,-1\}$. In this coordinate system, the unit normal vector is $n^{z}=-\frac{z}{\tl}$. Using the definition of the extrinsic curvature \eqref{K}, we obtain
\begin{align}
    K_{\mu\nu} &= \frac{\tl}{z^2}\tilde{h}_{\mu\nu}. \label{K-conformal}
\end{align}
Considering a metric perturbation, we assume it satisfies the FG gauge (i.e., $\delta g_{z\mu}=\delta g_{zz}=0$) and the TT gauge. The variation of the extrinsic curvature is then
\begin{align}
    \delta K_{\mu\nu} &= \frac{\tl}{z^2}\delta \tilde{h}_{\mu\nu} - \frac{\tl}{2z}\partial_{z}\delta \tilde{h}_{\mu\nu}. \label{variation of K}
\end{align}

\subsection{Extrinsic curvature of the entangling surface and extremal surface}\label{Kee section}
The induced metric on the extremal surface can be expressed in the ADM form:
\begin{align}
    d\hat{s}^2 &= N^2d\xi^2 + \sigma_{ab}(d\theta^a+N^a d\xi)(d\theta^b+N^b d\xi),
\end{align}
where $N=\frac{\tl T_0}{T_0^2-\xi^2}$, $\sigma_{ab}=\frac{N}{T_0}\tl \xi^2\theta_{ab},\xi^2=\tau^2+\textbf{x}^2,\textbf{x}\in\mathbb{R}^{d-2}$, and $\theta_{ab}$ represents the metric of the angular coordinates. In this background, the shift vector vanishes ($N^a=0$), and the covariant unit normal vector along the $\xi$-direction is $n_\xi=N=\frac{\tl T_0}{T_0^2-\xi^2}$.
The extrinsic curvature of the entangling surface is 
\begin{align}
    \k_{ab} &= \frac{1}{2}n^\xi\partial_\xi\sigma_{ab} \notag\\
    &= \frac{T_0}{\tl \xi}\sigma_{ab}.\label{K-entangling}
\end{align}
Now consider a perturbation of the metric:
\begin{align}
    d\hat{s}^2 &= (N^2+\delta\gamma_{\xi\xi})d\xi^2 + 2\delta \gamma_{\xi a}d\xi d\theta^a + (\sigma_{ab}+\delta \gamma_{ab})d\theta^a d\theta^b.
\end{align}
Using the definition of the extrinsic curvature, the variation is found to be
\begin{align}
    \delta \k_{ab} &= \frac{1}{2N}(\partial_\xi\delta \gamma_{ab} - \nabla_a\delta \gamma_{b\xi} - \nabla_b\delta \gamma_{a\xi}) - \frac{\delta \gamma_{\xi\xi}}{2N^2}\k_{ab}. \label{variation K pm}
\end{align}
To evaluate the extrinsic curvature of the extremal surface, we consider the bulk spacetime equipped with the metric
\begin{align}
    ds^2 &= \frac{\tl^2}{z^2}(-dx_{d-1}^2+dr^2+r^2d\theta^2+\xi^2d\Omega_{d-2}^2),
\end{align}
where $r^2=z^2+\xi^2$. As a consistency check, one can see that  $r^2d\theta^2\overset{r=T_0}{\to} \frac{T_0^2}{z^2}d\xi^2$ which gives the induced metric on the extremal surface.

The extremal surface is mathematically realized as a codimension-2 hypersurface. Its normal bundle is spanned by an orthonormal basis consisting of two spacelike/timelike unit normal vectors, denoted by $n_{A}^{\hat m}$ ($\hat m=1,2$). For this specific embedding, we identify these normals with the 1-forms along the $x_{d-1}$ and $r$ directions:
\begin{align}
    n^{(1)}_{x_{d-1}} &= \frac{\tl}{z},\\
    n^{(2)}_{r} &= \frac{\tl}{z}.
\end{align}
Defining the bulk coordinates as $x^{A}$ and the intrinsic surface coordinates as $y^{i}$, the surface is characterized by the embedding $x^{A}=X^{A}(y^i)$. The associated pushforward generators, $e^{A}_{i}=\frac{\partial X^{A}}{\partial y^i}$, span the tangent space of the surface. By adopting coordinates aligned with the embedding structure, $x^{A}=(x_{d-1},r,y^i)$, the tangent generators become trivial projections, $e^{A}_{i} = \delta^{A}_{i}$. 

The extrinsic curvature tensors corresponding to the two normal directions are constructed via the pullback of the covariant derivative:
\begin{align}
    \fk^{(1)}_{ij} &= e^{A}_i e^{B}_j \nabla_{A} n^{(1)}_{B} \notag\\
    &= 0,\\
    \fk^{(2)}_{ij} &= -e^{A}_i e^{B}_j \nabla_{A}n_{B}^{(2)} \notag\\
    &= 0.\label{Kfor extremal}
\end{align}
This result ($\fk^{\hat m}_{ij} = 0$) demonstrates that all components of the extrinsic curvature vanish identically.

\bibliographystyle{JHEP}
\bibliography{TE1_Law_in_Higher_Derivative_Gravity}

\end{document}